# Mars 2020 Perseverance rover studies of the Martian atmosphere over Jezero from pressure measurements

A. Sánchez-Lavega[1], T. del Rio-Gaztelurrutia[1], R. Hueso[1], M. de la Torre Juárez[2], G. M. Martínez[3], A.-M. Harri[4], M. Genzer[4], M. Hieta[4], J. Polkko[4], J. A. Rodríguez-Manfredi[5], M. T. Lemmon[6], J. Pla-García[5], D. Toledo[5], A. Vicente-Retortillo[5], Daniel Viúdez-Moreiras[5], A. Munguira[1], L. K. Tamppari[2], C. Newman[7], J. Gómez-Elvira[8], S. Guzewich[9], T. Bertrand[10], V. Apéstigue[8], I. Arruego[8], M. Wolff[11], D. Banfield[12], I. Jaakonaho[4], T. Mäkinen[4]

[1] UPV/EHU, Bilbao, Spain (e-mail lead author: agustin.sanchez@ehu.eus)
[2] Jet Propulsion Laboratory/California Institute of Technology, Pasadena, CA, USA
[3] Lunar and Planetary Institute, Houston, TX, USA
[4] Finnish Meteorological Institute, Helsinki, Finland
[5] Centro de Astrobiología (INTA-CSIC), Madrid, Spain
[6] Space Science Institute, Boulder, CO 80301, USA
[7] Aeolis Research, Pasadena, CA, USA
[8] Instituto Nacional de Técnica Aeroespacial, INTA, Madrid, Spain
[9] NASA Goddard Space Flight Center, Greenbelt, MD, USA
[10] Observatoire Paris Meudon, Paris, France
[11] Space Science Institute, Brookfield, WI, USA
[12] Cornell University, NY, USA

Corresponding author: first and last name (agustin.sanchez@ehu.eus)

**Key Points:**

- We study the pressure measurements performed on the first 460 sols by the rover Perseverance M2020
- The daily and seasonal cycles and the evolution of six tidal components and their relationship to dust content are presented.
- We characterize long-period waves (sols), short-period gravity waves (min.), rapid pressure fluctuations and a regional dust storm impact.

**Abstract**

The pressure sensors on Mars rover Perseverance measure the pressure field in the Jezero crater on regular hourly basis starting in sol 15 after landing. The present study extends up to sol 460 encompassing the range of solar longitudes from $L_s \sim 13°$ - $241°$ (Martian Year (MY) 36). The data show the changing daily pressure cycle, the sol-to-sol seasonal evolution of the mean pressure field driven by the $CO_2$ sublimation and deposition cycle at the poles, the characterization of up to six components of the atmospheric tides and their relationship to dust content in the atmosphere. They also show the presence of wave disturbances with periods 2-5 sols, exploring their baroclinic nature, short period oscillations (mainly at night-time) in the range 8-24 minutes that we interpret as internal gravity waves, transient pressure drops with duration $\sim$ 1-150 s produced by vortices, and rapid turbulent fluctuations. We also analyze the effects on pressure measurements produced by a regional dust storm over Jezero at $L_s \sim 155°$.

**Plain Language Summary**

Mars rover Perseverance landed on 18 February 2021 on Jezero crater. It carries a weather station that has measured, among other quantities, surface atmospheric pressure. This study covers the first 460 sols or Martian days, a period that comprises a large part of the Martian year, including spring, summer and a part of autumn. Each sol, the pressure has significant changes, and those can be understood as a result of the so-called thermal tides, oscillations of pressure with periods that are fractions of one sol. The mean value of pressure each sols changes with the season, driven by the $CO_2$ sublimation in summer and condensation in winter at both poles. We report oscillations of the mean daily pressure with periods of a few sols, related to waves at distant parts of the planet. Within single sols, we find oscillations of night pressure with periods of tens of minutes, caused by gravity waves. Looking at shorter time intervals, we find the signature of the close passage of vortices such as dust devils, and very rapid daytime turbulent fluctuations. We finally analyze the effects on all these phenomena produced by a regional dust storm that evolved over Jezero in early January 2022.

**1 Introduction**

Perseverance rover landed on Mars on 18 February 2021 at Jezero crater at longitude 77.45°E and latitude 18.44°N (Newman et al., 2022). Onboard is the *Mars Environmental Dynamics Analyzer* instrument (MEDA) a suite of sensors dedicated to study the atmospheric dynamics and aerosol content and properties (dust and clouds) (Rodríguez-Manfredi et al., 2021a). Among them is the Pressure Sensor (PS) whose properties are described in Rodríguez-Manfredi et al. (2021a) and whose operational details are given in a companion paper in this issue (Harri et al., 2022). The MEDA PS is a device based on the silicon-micro-machined capacitive pressure sensor head (Barocap®) and transducer technology developed by Vaisala Inc (Finland). Pressure moves the capacitor plates in the sensor, changing its capacitance. The technology of the Barocap© has previously own in 6 missions and MEDA PS design is very similar to REMS-P (Rover Environmental Monitoring Station-Pressure) (Harri et al., 2014). The PS is located inside the MEDA Instrument Control Unit (ICU) in the rover body, with a filter-protected tube connecting it to the ambient atmosphere. The absolute accuracy is < 3.5 Pa and the electronic noise is $\sim$ 0.13 Pa. The pressure is measured with a cadence of 1 Hz and for one-hour periods typically separated by one-hour-long periods, sampling each local time every two consecutive sols. Data are available in Rodríguez-Manfredi et al. (2021b). We use for the temporal representation the Martian Local Mean Solar Time (LMST). In this paper we focus on the analysis of the measurements provided by the PS and on the study of the different dynamical mechanisms that intervene in the pressure field at the different spatial and temporal scales. The studied period begins in sol 15 (first PS data) corresponding to Mars solar orbital longitude $L_s = 13°$ (following Northern hemisphere spring equinox) and ends at sol 460 corresponding to $L_s = 241°$ in the autumn season (Northern hemisphere autumn equinox is at $L_s =180°$) before reaching the perihelion ($L_s = 251°$) of MY 36.

The PS data gathered by Perseverance in Jezero allows us to study the combined action of periodic, non-periodic and transient dynamical mechanisms. Among them, the seasonal evolution due to deposition and sublimation of the $CO_2$ in the polar caps, the thermal tides, waves of periods above 1 sol, short period waves (in the range of minutes), mesoscale dynamics, passage of transient vortices and dust devils, and high frequency oscillations due to convection and turbulence, as well as transient instabilities and a regional dust storm around sol 313, modify strongly the pressure field. For a review of previous studies of all these phenomena see Read et al (2015) and Barnes et al. (2017).

Figure 1 shows the coverage and evolution of daily pressure values with good temporal coverage starting on sol 15 and ending in sol 460. It shows, at a quick glance, the daily and the seasonal cycle in pressure, as well as specific deviations from the general trends that we will describe in detail below.

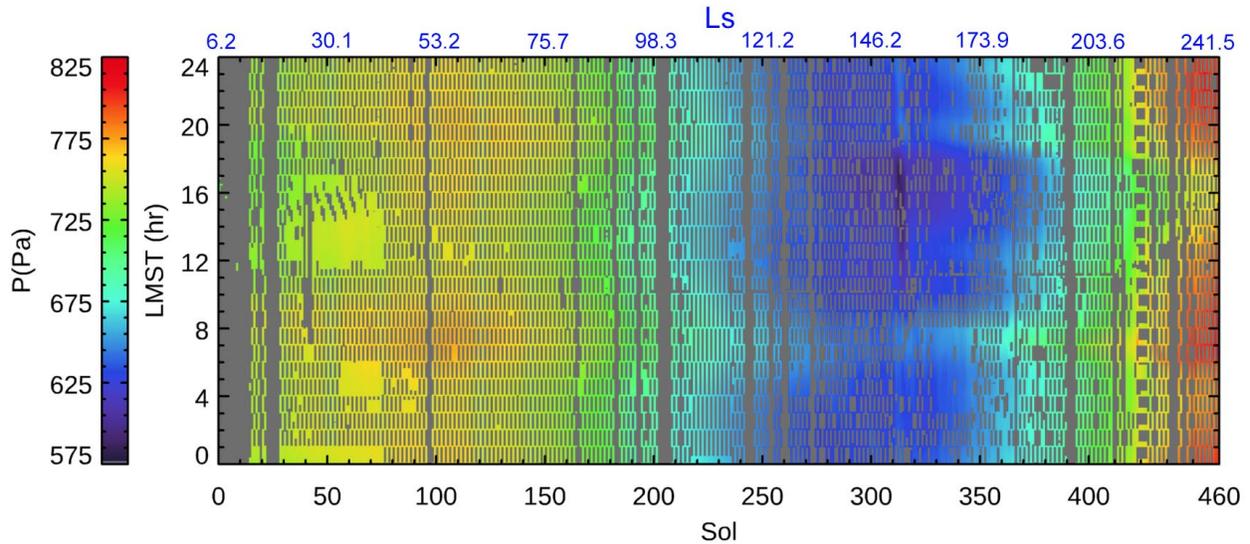

**Figure 1.** *A synthesis of the daily surface pressure measurements coverage (vertical axis LMST, Local Mean Solar Time) as a function of the sol number and solar longitude of Perseverance rover on Mars. The pressure is given in Pascal (Pa). Data are available in Rodríguez-Manfredi et al. (2021b).*

## 2. The daily pressure cycle

The daily pressure cycle at Jezero showed a rich morphology and variability during the first 460 sols of the mission, in response to the presence of different dynamical mechanisms in the atmosphere that act at multiple spatial and temporal scales. The pressure tidal components dominate the daily variation of pressure. Figure 2 shows examples of the daily pressure cycle at selected sols based on $L_s$ interval length to evenly cover the whole range of $L_s$ analyzed. In general, there is a maximum in pressure at about 7-8 hr Local Mean Solar Time (LMST), with a secondary maximum close to 19-23 hr LMST (sometimes becoming the absolute maximum, Figure 2), and a minimum (sometimes double) that shifts with season between 12 hr and 16 hr LMST. During the insolation hours (broadly speaking from 6 to 18 hr LMST, Munguira et al., 2022), the atmosphere becomes convectively unstable, vortices and dust devils develop and temperature fluctuations and thermal turbulence intensifies (Read et al., 2017). On the contrary, during nighttime, the atmosphere becomes thermally stable and pressure oscillations with periods of minutes develop (see section 6). The comparison of these results with model predictions (Newman et al., 2021; Pla-García et al., 2021) is presented in detail in Harri et al. (2022; this issue), where it is shown that the general behavior of the daily cycle of pressure fits reasonably well with the predictions of the daily maxima and minima, although there are differences in the secondary minimum and maximum. These are due to the combination of the different contributions of the tidal components in response to the aerosol distribution. In Supplementary Figure S1 we compare the above daily pressure cycles in Figure 2 with the model predictions by the Mars Climate Database (MCD) for the case "climatology average solar" (Forget et al., 1999; Millour et al., 2015). The global trend is captured but part of the differences could be due to the low-resolution of the model as compared to the local measurements. Regional and local circulation effects in the Jezero crater are expected to be less than 1%-3% (Newman et al., 2021; Pla-García et al., 2021). Unlike Gale crater where slope flows play a major role in the dynamics (Richardson and Newman, 2018) and observed pressure variations are 13% (Haberle et al., 2014), Jezero is a shallow crater with no central peak, and therefore the effect of internal circulation on the daily cycle is expected to be weaker (Tyler and Barnes, 2015).

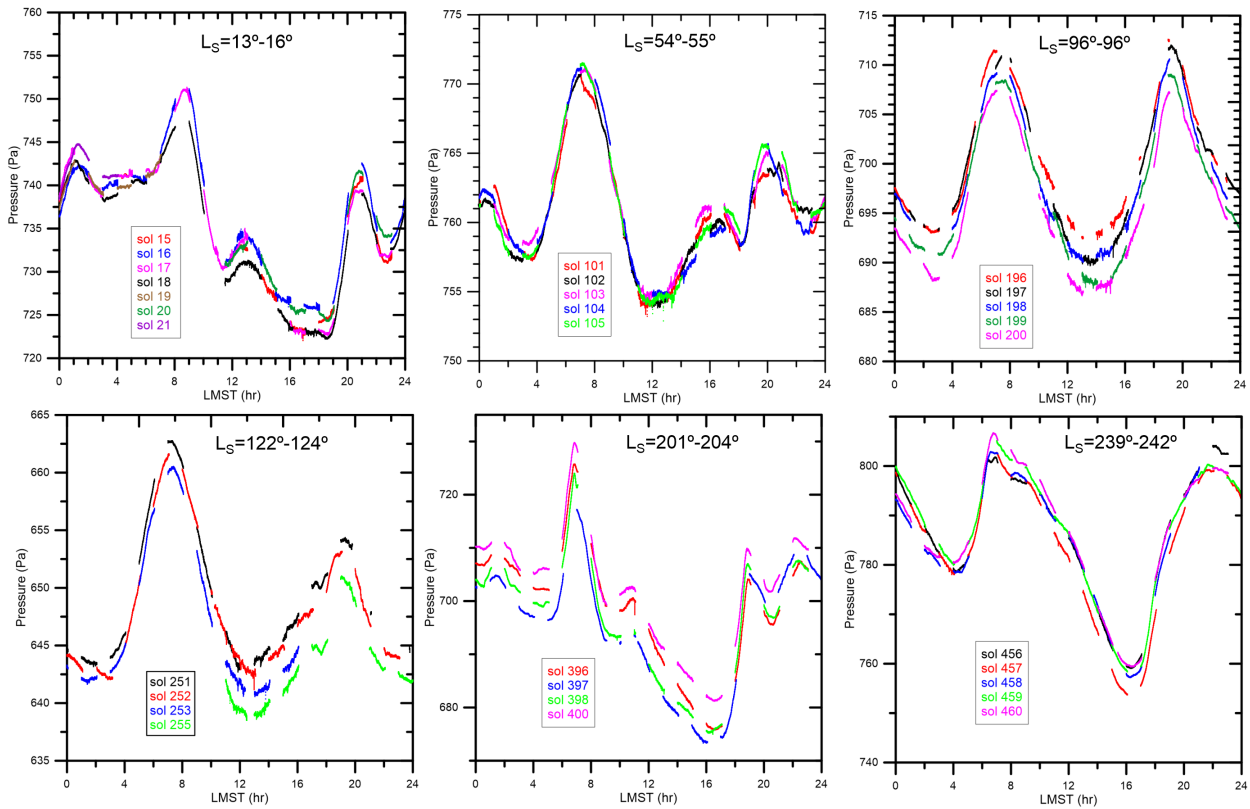

**Figure 2**. *Examples of the daily pressure cycle along the studied period grouped in sets of five sols to visualize the changes over short (minute-hours) and long periods (several sols) of time. The corresponding average solar longitudes are given as insets. Note that the pressure-scale is different in each panel. For a comparison with MCD predictions see Supplementary Figure S1.*

## 3. The seasonal cycle

It is well known since the first in-situ measurements of pressure by Viking landers (Hess et al., 1977, 1980; see Martínez et al., 2017 for a comparative review), that the yearly evolution of the pressure field is dominated by the sublimation and condensation of $CO_2$ over the poles and the associated atmospheric mass transport (Wood & Paige, 1992; Khare et al., 2017). Perseverance first PS measurements were taken at $L_s \sim 15°$, in the early northern spring-time season (Figure 3). The mean daily pressure initially increased from 735 Pa on sols 15-20 ($L_s = 13°$-$16°$) to a maximum of 761 Pa around sol 105 ($L_s = 55°$) in northern spring. Then it gradually decreased to 625 Pa in sol 306 ($L_s = 101°$) during the southern winter. Note however that shortly after this minimum the mean pressure showed an abrupt decrease around sols 311-318 due to the evolution of a regional dust storm over Jezero crater (section 8). Following this minimum, the pressure trend increased with the arrival of the southern spring season ($L_s = 180°$) until sol 460 which is the last measurement reported in this paper.

The dispersion in the pressure measurements reflects both the variability of the mean pressure field due to the different atmospheric phenomena here discussed and the dispersion resulting from the number of measurements used each sol. We excluded for this analysis sols with incomplete measurements along large sectors of the daily cycle (see Figure 1) and sols where the overall number of measurements is below $3 \times 10^4$. The average data points coverage is $51600 \pm 10600$, with a maximum of $8.45 \times 10^4$. The mean standard deviation (std) in the daily mean pressure measurement is 7.3 with a std of 2.9 Pa. The std of the mean daily pressure ranged from 4 Pa to 18.7 Pa when the dust storm reached over Jezero.

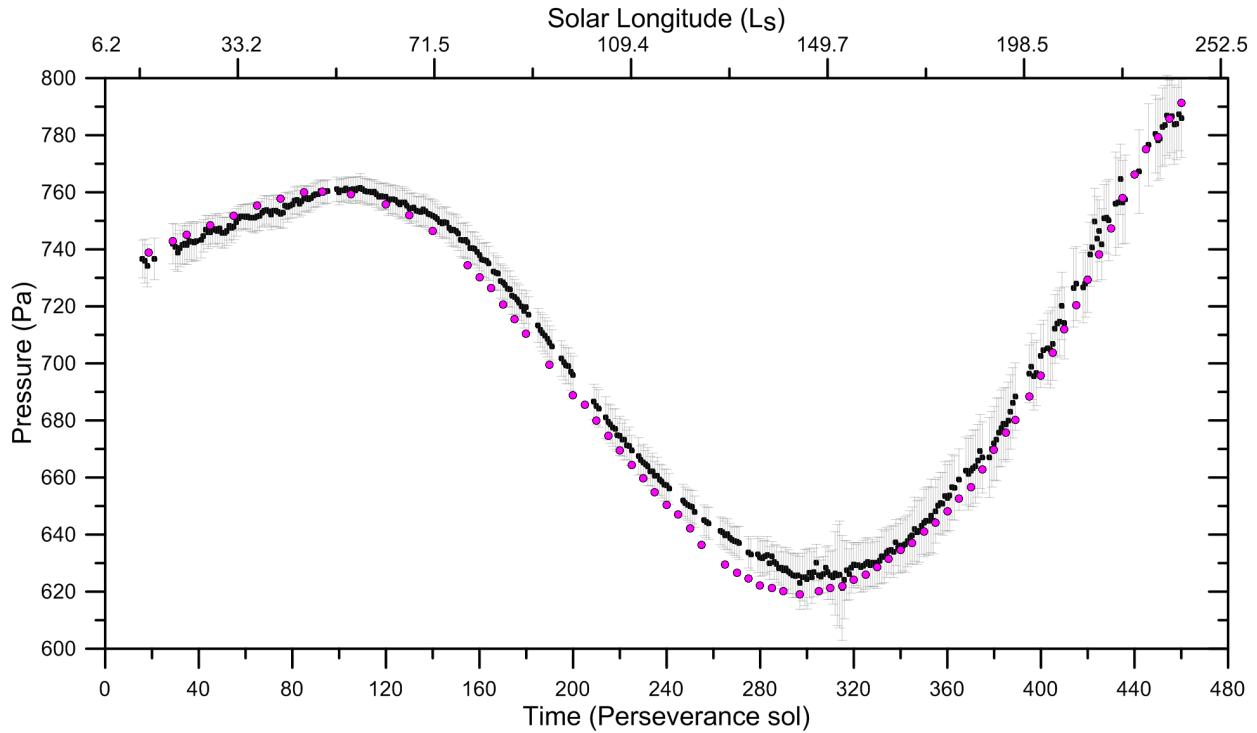

**Figure 3**. *Seasonal evolution of the mean sol pressure between sols 15 and 460 ($L_s = 13°$ to $L_s = 241°$). Black dots and gray bars represent the mean pressure and its standard deviation for each sol. The magenta dots are predictions by the MCD on a nominal scenario.*

Figure 3 also includes the mean daily pressure predicted by the MCD (Forget et al., 1999; Millour et al., 2015). Typically, the differences between the observed and predicted pressure values are ~ 4-6 Pa but a large deviation is found between sols 250-300 where differences reach 10 Pa. This corresponds to a period with variability in the atmospheric aerosol content (dust and water ice clouds) that is probably not captured in detail by this model. In addition, the comparison of the seasonal trend between several models (Newman et al., 2021, their figure 5; Harri et al., 2022) predict the minimum at $L_s \sim 145°$ (observed $L_s \sim 150°$) and differences between different models between 20 and 30 Pa. On average, the models fit globally the observed seasonal behavior, but each model performs better in different ranges of solar longitudes.

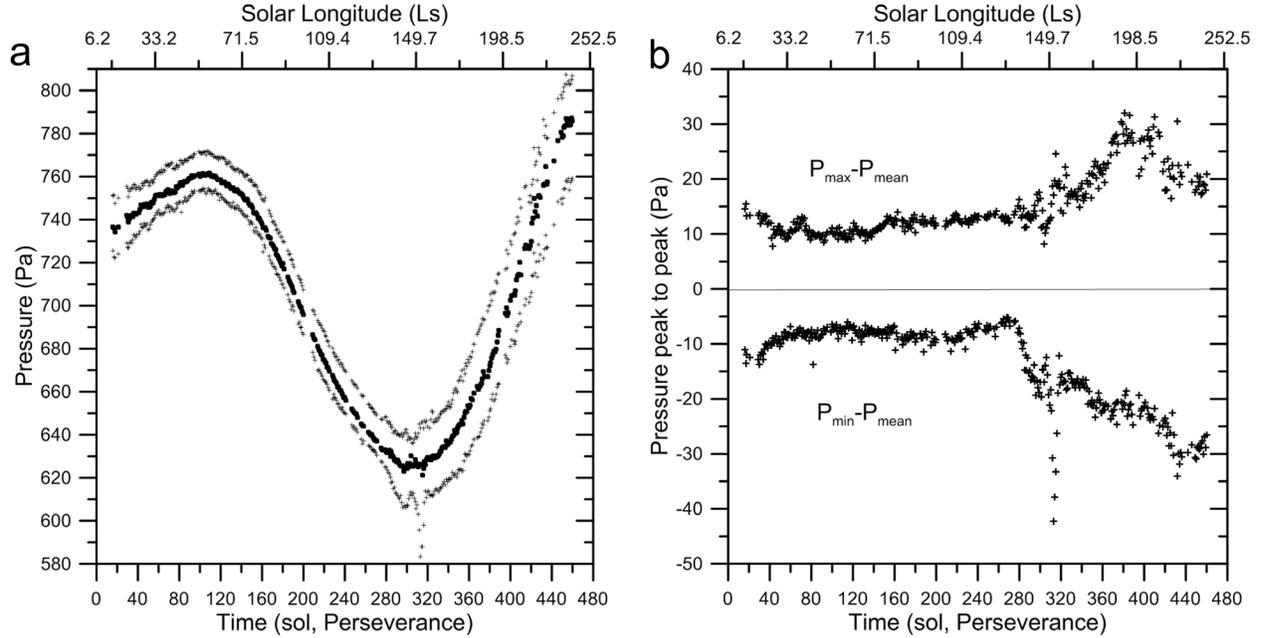

**Figure 4**. *(a) Seasonal evolution of the minimum and maximum (crosses) and mean (dots) pressure values for each sol; (b) Differences of maximum and minimum pressures relative to their mean value for each sol.*

To show more clearly the range of daily pressure variability we present in Figure 4 the minimum and maximum pressures for each sol, and their differences with the daily mean value. Up to sol 275 the peak to peak differences were ~ 20 Pa and then increased to maximum values of ~ 50 Pa as a consequence to the increase of the aerosol content in the atmosphere. The figure also clearly shows the arrival and transit of the dust storm around sol 313, which produced a sharp drop in the pressure minimum, as will be discussed in section 8.

## 4. Thermal Tides in the Pressure field

The main atmospheric dynamical phenomena controlling the daily pressure cycle are the thermal tides (Zurek, 1976; Hess 1977; Read et al., 2015; Guzewich et al., 2016; Barnes et al., 2017). These are planetary atmospheric oscillation modes forced by solar heating with periods that are harmonics of the solar day. The tides redistribute mass and so manifest at the surface in the temperature, wind and pressure fields, as described by the classical tidal theory (Chapman and Lindzen, 1970; Wilson, and Hamilton, 1996). To capture the evolution of the amplitude and phase of the tidal components we made a Fourier analysis of the daily pressure cycle (Figure 2) using a discrete Fourier Transform of the pressure measurements for series with more than 30,000 data points. In a second study we performed the Fourier analysis on the combination in a single daily cycle of the data corresponding to two consecutive sols to fill data gaps. The Fourier series is expressed in terms of sines and cosines with Fourier coefficients ($a_n, b_n$) respectively. We present the amplitude $S_n = \sqrt{a_n^2 + b_n^2}$ and phase $\varphi_n = \arctan(b_n / a_n)$ of the first six components, $n = 1$ (diurnal, 24 hr period Martian time), 2 (semidiurnal, 12 hr), 3 (terdiurnal, 8 hr) etc. The phase $\varphi_n$ represent, when scaled into hours in the time interval of 24 hr, the local time of maximum pressure for each term of the harmonic series. $S_0$ is approximately the mean pressure value for each particular sol (= $P_{mean}$). We present in Figure 5 the seasonal evolution of the amplitude of the first six harmonics for which the pressure signal is well above the noise level. Both Fourier analyses, the single sol and the two consecutive sols, are in very good agreement and follow a similar seasonal trend for all the tidal components. In Supplementary Figure S2 we show examples of the Fourier fits and the resulting residuals compared to the measurements. The fit to 12 Fourier terms of the daily pressure cycle generates residuals < 0.2 Pa, below the amplitude of the tidal components (Figure 5).The analysis shows that there is also a contribution of higher order harmonics (7 and 8) in specific sectors of the seasonal evolution of the tidal components. We hope to work on this point in a future analysis when the data for a full Martian year are available. These first six modes ($n$ = 1-6) show maximum amplitudes in the range from 2 to 16 Pa, although a transient intense peak above the mean values can be distinguished in the diurnal and semidiurnal component at sol 313 due to the presence of the regional dust storm (Figure 5). Significant is the high anti-correlation between amplitudes $S_1$ and $S_2$ from the beginning of the mission to the sol ~ 330 ($L_s$ ~ 175°), the simultaneous increase in $S_4$ and $S_6$ with the dust season (since ~ sol 280, $L_s$ ~ 136°) with a peak of 4 Pa, and the similar trends in $S_3$ and $S_5$.

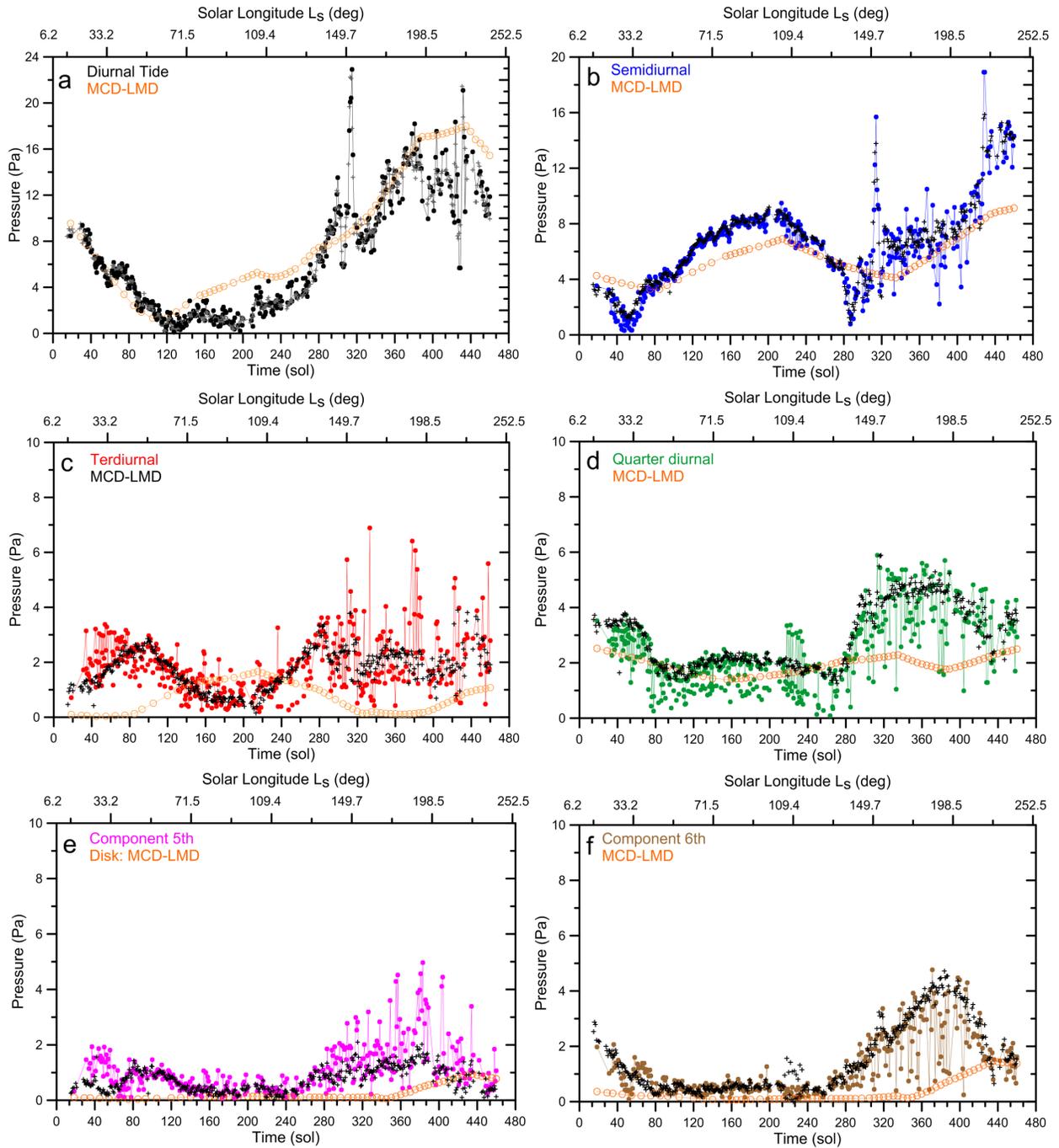

**Figure 5.** *Evolution of the amplitude of the first six tidal modes along the studied period given in sols and in solar longitude). Each tidal component is identified by its name and color for a Fourier analysis of a single sol. The gray crosses in the diurnal mode (a) and the black crosses in the other modes (b-f) are Fourier analysis of a diurnal cycle combining two consecutive sols. The brown color disks show the prediction by the MCD model.*

The measured amplitudes are compared with those calculated using the daily pressure cycle prediction by the MCD model. The comparison shows that both the diurnal and semidiurnal tides follow reasonably well the predicted trends. This is also the case for the $4^{th}$-$5^{th}$-$6^{th}$ components. However, the terdiurnal tide follows a trend opposite to that predicted by the model. The diurnal tide is sensitive to forcing by heating in the lower atmosphere (and so to dust content), interacting with topography and with other spatially variable surface factors, driving motions vertically with a wavelength of ~ 35 km (Zurek, 1976; Read et al., 2015). The semidiurnal tide is also sensitive to dust heating but in a much larger vertical scale, with wavelengths ~ 100 km (Read et al., 2015). Changes in the dust content and in water ice clouds (in particular in the aphelion season) and in their vertical distribution, could be behind the observed differences with the model for most of the components.

Figure 6 shows the phase evolution of the first three tidal components. The phase of the diurnal component is normally ~ 4 hr (3-8 hr range), but it underwent an abrupt jump of about 8 hr between sols 120 and 210 (around $L_s$ ~ 90°). The phase of the semidiurnal component was ~ 3-6 hr in the first 270 sols of the mission The phase of the semidiurnal component was 3-6 hr in the first 270 sols of the mission ($L_s$~6.2°-132°), but showed a sharp drop at around sol 285 ($L_s$~139°), followed by a rapid shift towards later hours coincident with the increase in optical opacity toward sol 290 ($L_s$~142°). This shift continued until the arrival of the dust storm on sol 310 as will be described in more detail in section 8.

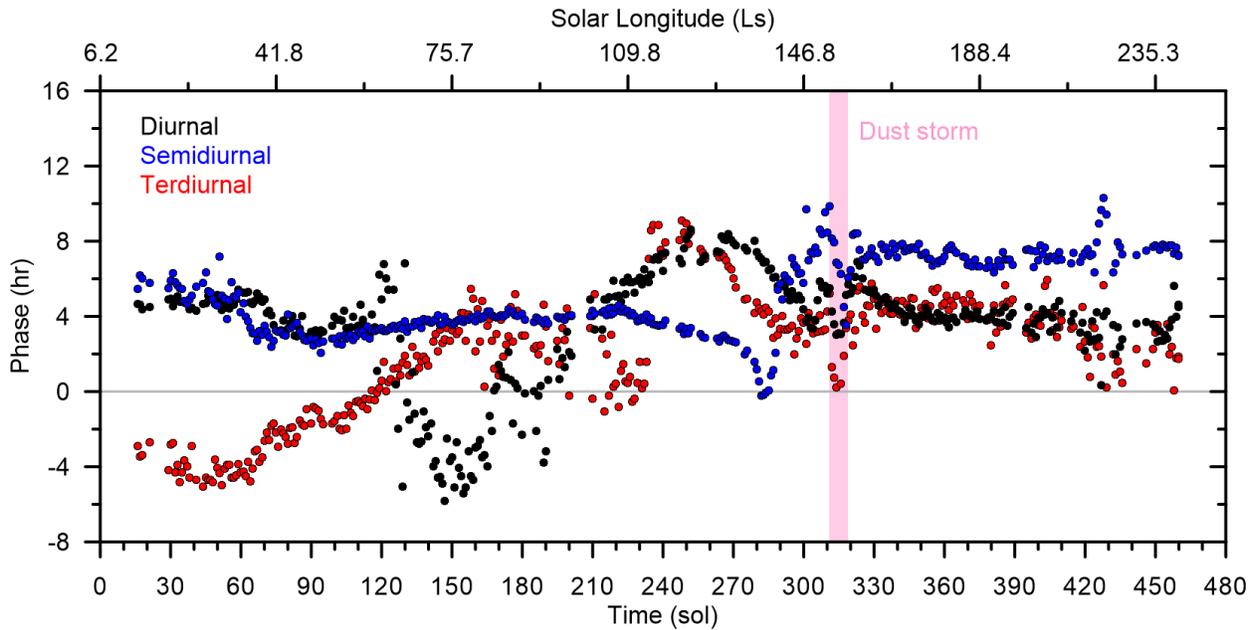

**Figure 6.** *Evolution of the phase of the three tidal modes (diurnal, semidiurnal, terdiurnal) along the studied period given in sols and in solar longitude). Each tidal component is identified by its name and a color given in the inset. The vertical pink bar corresponds to the period when the dust storm developed over Jezero.*

The normalized amplitude ($\delta P/P_0$) of the diurnal and semidiurnal tides showed ample variability in the studied period (Figure 7). Out of the stormy sols, the maximum observed relative change is in the diurnal tide with $(\delta P_1/P_0)_{max}$ ~ 0.025 followed by the semidiurnal with $(\delta P_2/P_0)_{max}$ ~ 0.020.

Guzewich et al. (2016) studied the behavior of the first four tidal components during more than a Martian year from Mars Science Laboratory (MSL) REMS pressure measurements at Gale crater. Outside dust storms, for the same period of $L_s$ studied here, they found higher values for $S_1$ with $(\delta P_1/P_0)_{max} \sim 0.05$ but similar for $S_2 \sim 0.02$, and phases in the same range of hours but with a different behavior. The differences are probably related to the different latitude of both rovers and crater topography and circulation in the case of Gale crater.

The studies of Zurek and Leovy (1981) from Viking data and more recently of Guzewich et al. (2016) and Ordóñez-Etxeberría et al. (2019) from MSL showed clearly the link between the atmospheric opacity and diurnal and semidiurnal tide amplitudes. MSL studies showed both tides to be correlated at a 90% level, with a very strong relationship between atmospheric aerosol loading and tide amplitude. The terdiurnal and quarter diurnal have however almost zero correlation and a modest anti-correlation, respectively. Wilson et al. (2008) found that the amplitude of the semidiurnal tide ($S_2$) is directly related to the vertically integrated optical depth $\tau$ (dust and clouds) in the atmosphere and proposed the empirical relation $(\delta P_2/P_0)(\%) = 1.6\tau + 0.3$ (see also Barnes et al., 2017). Since the visible optical depth has been measured at Jezero crater on a daily basis (although at different LMSTs) from images obtained by Perseverance cameras Mastcam-Z and Skycam on MEDA instruments (Lemmon et al., 2022), we have used these data to plot the above function in Figure 7. Even though there is in general a reasonable agreement with our $S_2$ retrieval, this is not the case during the first part of the mission (from the first sols and up to sol ~ 125). In particular, the semidiurnal component showed a strong drop in amplitude between sols 20-75 that is not related to the aerosol content. This particular issue deserves a more detailed study, which is currently under way, but that is beyond the scope of this paper.

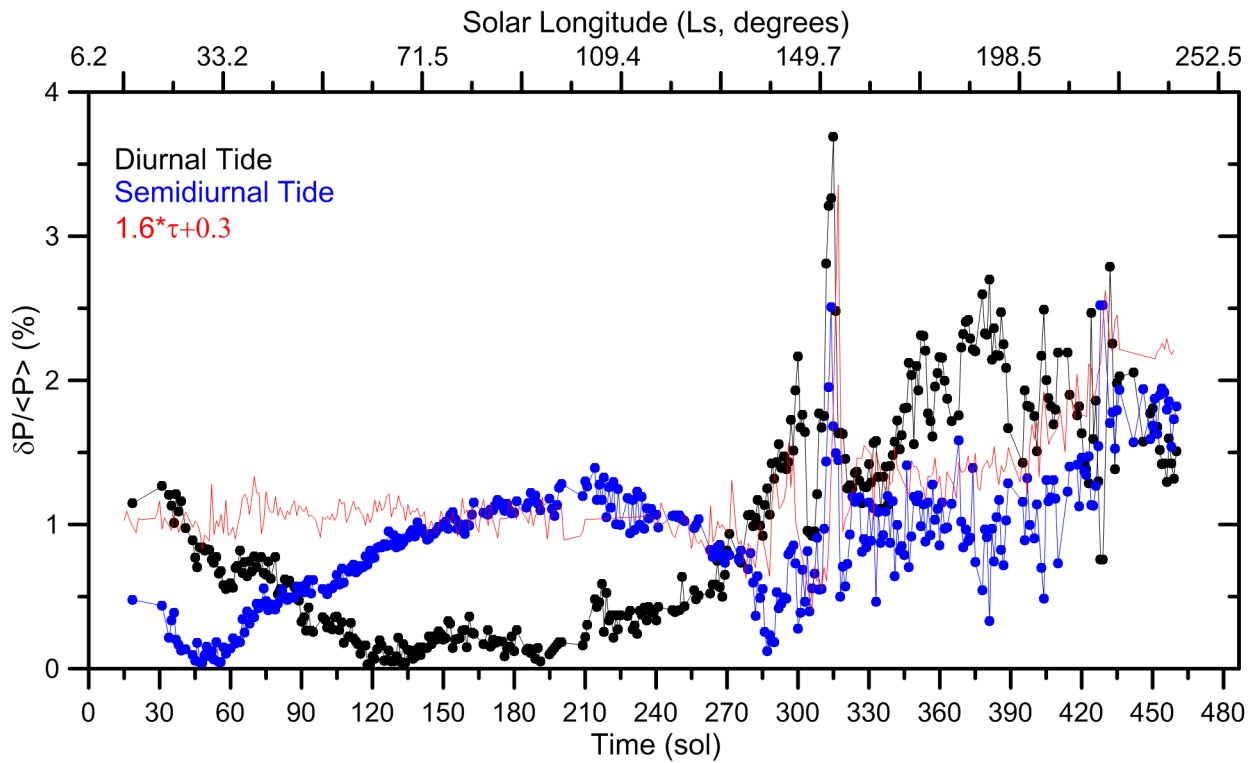

**Figure 7**. *Normalized amplitudes of the diurnal (black dots) and semidiurnal (blue dots) components of the thermal tide compared with the scaled optical depth ($\tau$) given by the relation $1.6\tau + 0.3$ (thin red line).*

In order to further specify a relationship between the amplitude of the tidal components and the dust content, we show in Figure 8 different linear combinations of the measured amplitudes of the tidal components $\sum_n c_n S_n \approx c_1 S_1 + c_2 S_2 + c_3 S_3 + c_4 S_4$ up to the 4$^{th}$ mode (where $c_n$ are constant coefficients), and compare them with the measured optical depth $\tau$ (scaled). The first and simplest approach is to see if a 50% linear combination of the diurnal and semi-diurnal components correlates with the evolution of the dust content in the atmosphere. It can be seen that $\tau$ correlates reasonably well for such case ($c_1 = c_2 = 0.5$), including the dust storm event, but that to fit the dusty period after the storm, we need to include the components $S_3$ and $S_4$. A good correlation is found with coefficients $c_1 = 0.27$, $c_2 = 0.53$, $c_3 = 0.16$, $c_4 = 0.03$, which minimizes the quadratic distance between the predicted value and scaled $\tau$, using the Nelder Mead algorithm as implemented in scipy.optimize library (https://docs.scipy.org/doc/scipy/tutorial/optimize.html; Nelder and Mead, 1966). This is just an empirical search to try to see the action of dust on the different components of the tide. Our conclusion is that in the dusty epoch (starting after sol ~ 320), the aerosol content affects differentially the first four modes of the tides, and in a different manner during the dust storm, which only affects $S_1$ and $S_2$.

The increase in $S_5$ and $S_6$ after sol 280 correlates with the opacity increase in Jezero, a tendency also shown by the model predictions (Figure 5e-f). The inverse behavior observed in $S_3$ in relation to the model prediction remains to be explained.

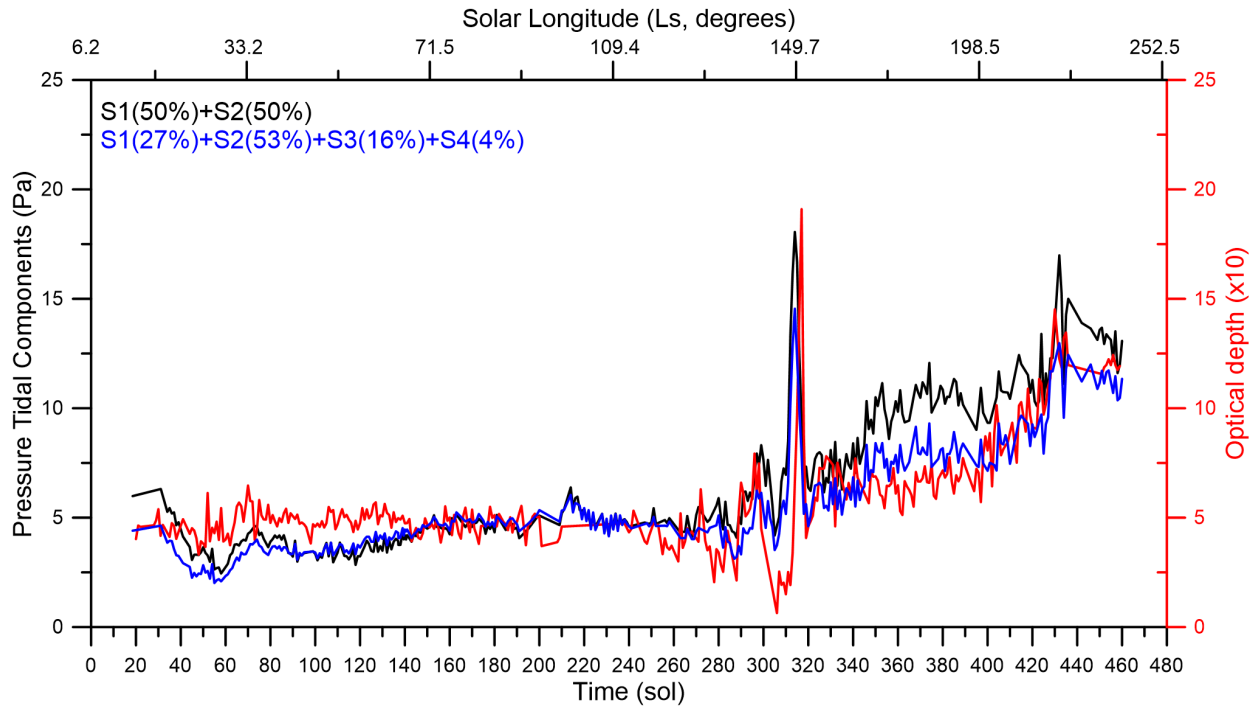

**Figure 8**. *Comparison between the optical depth evolution (x10, red line) and two different linear combinations of the diurnal and semidiurnal tidal components (black line) and the first four components in the proportion indicated in the figure inset (blue line).*

## 5. Long-period (2-10 sol) oscillations: baroclinic waves

Looking at the seasonal evolution of mean pressure, the presence of oscillations around the mean trend appears evident (see for example Figures 3 and 4 between sols 15-105). In order to study the amplitude and period of these oscillations, we initially performed polynomial fits to the seasonal curve up to degree 11. However, fits are improved when considering shorter periods and therefore we divided the curve into different sections, fit each sector, and finally combined the results for the whole period as shown in Figure 9. The sols considered in the fits were as follows (i is the polynomial degree): 16-105 (i=5), 105-280 (i=6), 279-330 (i=2), 330-405 (i= 3), 395-460 (i=3). In all these cases the coefficient of determination (r-squared) > 0.98 for polynomial terms with degree greater than zero. Tests performed with fits to other longer time periods of sols, for example for 209-405 (i=8), gave similar results.

The de-trended data show oscillations dominated by a mean period in the range 2 - 5 sols, resulting from the average of peak-to-peak times, which vary within a broad range of periods from ~ 3 to 10 sol. Longer periods, 2 or 3 times this value, cannot be ruled out, but their detailed study and characterization is beyond the scope of this work and will be presented elsewhere. The amplitude *A* obtained from the residuals relative to the mean fitted value changed notoriously in time. We found a mean value *<A>* = 1.8 Pa (peak to peak 3.6 Pa) between ~ sols 16 – 80, decreasing to 0.9 Pa in sols ~ 90 – 280, and increasing again to 2.7 Pa when the optical depth started to grow on sol 280 and up to sol 375 (including an abrupt change during the dust storm described in section 8). The oscillations become more pronounced in amplitude reaching ~ 6 Pa during the period of highest optical opacity (see Figure 8) ranging from sol ~ 375 to the last one analyzed, sol 460 (Ls ~ 200°-240°). The amplitude of the oscillations has certain correlation with the amount of dust in suspension in the atmosphere. The degree of such correlation is however weak, as shown in Supplementary Figure S3.

Similar pressure oscillations have been detected in past missions, with pressure sensors showing oscillations in the surface pressure with typical periods of 2–8 sol and amplitudes of a few percent but varying in intensity depending on location on Mars. For example, oscillations were much larger at 48°N (Viking Lander 2) than at 22°N (Lander 1) (Barnes 1980, 1981, 1984) and produced a discernible effect at latitudes close to the equator, as at 4.5°S in Gale Crater (Haberle et al., 2018) and at 4.5°N in Elysium Planitia (Banfield et al., 2020). These oscillations have been interpreted as the signature of high frequency travelling waves arising from baroclinic instabilities in mid and high northern latitudes (Leovy 1979; Barnes 1980, 1981, 1984; Tyler and Barnes, 2005; Hinson and Wang, 2010; Barnes et al., 2017; Haberle et al., 2018; Banfield et al., 2020). These disturbances manifest sometimes in images taken by orbital vehicles as dust storms and cloud systems with a variety of morphologies (spirals, textured storms, arc-flushing storms) and are primarily confined to the polar cap edge and mid-northern latitudes during the autumn, winter, and spring seasons (James et al., 1999; Cantor et al., 2001; Wang et al., 2003, 2005; Khare et al, 2017).

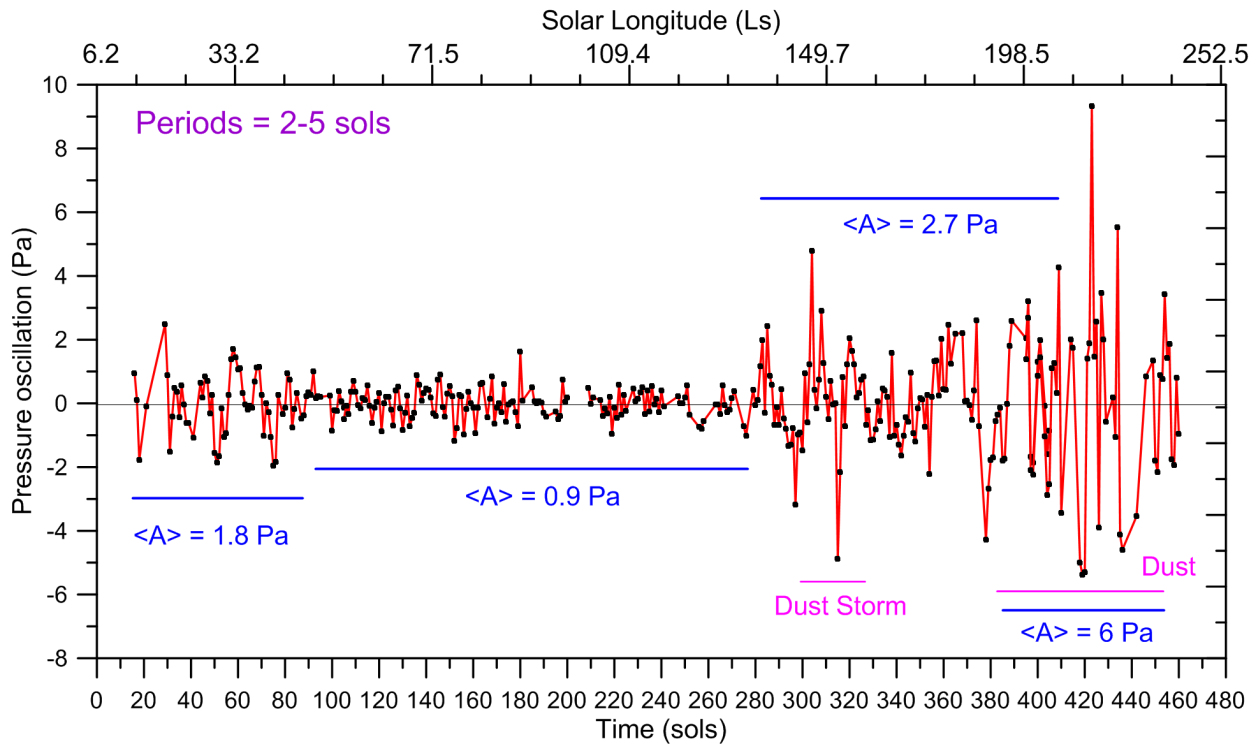

**Figure 9**. *Pressure oscillations obtained as residuals between the measured mean daily pressure and polynomial fits to the seasonal evolution curve. The de-trended data to the seasonal trend have been divided in four temporal sectors according to their different mean pressure amplitude <A> (in Pascal).*

The disturbances grow from the baroclinic and barotropic instabilities in the eastward jet at the edge of the North Polar Cap edge in high northern latitudes ~ 60°N-80°N (Barnes, 1984; Barnes et al., 2017). This jet shows intense vertical wind shear following the thermal wind balance and according to the north-south temperature gradient as predicted by General Circulation Models (GCM). We show in figure 10 the variability of this jet in its intensity and altitude-latitude location at Jezero longitude during the studied period according to the MCD model (Forget et al., 1999; Millour et al., 2015). It is reasonable to assume that the high temporal variability of the jet is behind the changes observed in the wave activity shown in figure 9. We note that the barotropic portion of the waves is expected to increase as time approaches winter solstice ($L_s=270°$).

To characterize the horizontal scale of the eddies produced by the baroclinic instability we first estimate the Rossby deformation radius (Vallis, 2006) defined as $L_D = NH/f$, where $f = 2\Omega \sin\varphi = 1.22 \times 10^{-4}$ s$^{-1}$ at $\varphi=60°$N, $N$ is the Brunt-Väisälä frequency $N^2 = (dT/dz + g/C_P)g/T$ and $H = R^*T/g$ the scale-height (In these equations, $\Omega$ is Mars angular velocity, $\varphi$ the latitude, $g$ acceleration of gravity and $C_P$ and $R^*$ the constant pressure capacity and gas constant of Martian atmosphere). We focus on the springtime period ($L_s$ ~ 30°-60°) where the storm activity is particularly high at the North Polar Cap (NPC) edge (Read et al., 2015; Clancy et al., 2017). Analysis of the images of the disturbances indicate that the dust and clouds extend vertically from the surface to h ~ 10 km (Sánchez-Lavega et al., 2018, 2022). Then, using $C_p$ = 780 Jkg$^{-1}$K$^{-1}$, $R^*$ = 192 J kg$^{-1}$K$^{-1}$, $g$ = 3.72 ms$^{-2}$ and taking an average temperature $T$ ~ 193 K and dT/dz ~ 2x10$^{-3}$ km m$^{-1}$ in this altitude range at latitude 60°N from the MCD, we find that $N$ ~ 0.012 s$^{-1}$, $H$ = 10 km and $L_D$ ~ 1000 km.

For waves with equal zonal and meridional wavenumbers ($k = \ell$), the maximum growth rate corresponds to $L_{Bclin}$ ~ (3.9-5.5) $L_D$ ~ 3900 - 5500 km (Lin, 2007; Vallis, 2006) and the wavenumber of the disturbances is $n = 2\pi R_M \cos\varphi / L_{Bclin}$ ~ 2-3 ($R_M$ = 3389 km). The phase speed of these waves is given by $c_x = (H/2)(\partial u/\partial z)$ ~ 20 ms$^{-1}$ where we used $(\partial u/\partial z)$ ~ 0.004 s$^{-1}$ from figure 10 in the altitude range 0–15 km. This is probably and upper phase speed limit. Then, the corresponding translation zonal velocity relative to the mean flow is $U_{disturbance} = U_0 - c_x$. Since $U_{disturbance}$ (observed) = $L_{Bclin}/\tau$ ~ 10 ms$^{-1}$, we deduce a background wind speed $U_0$ ~ 30 ms$^{-1}$. These numbers agree with what is usually measured for these disturbances (Sánchez-Lavega et al., 2018, 2022).

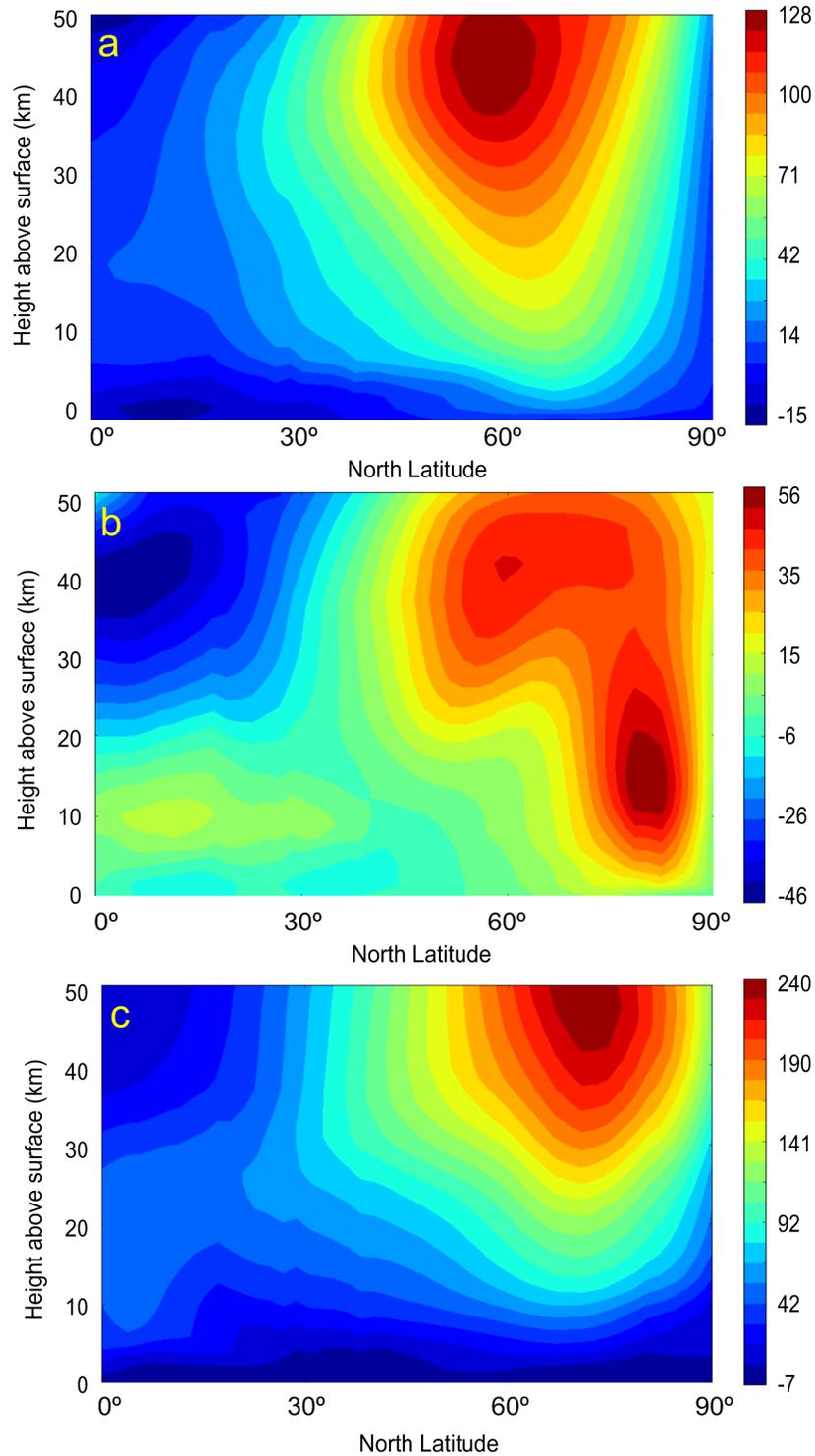

**Figure 10.** *Zonal wind velocity maps in altitude (0-50 km) and latitude (Equator to Pole) at 12 hr local time for the following dates: (a) $L_s=15°$ (sol 19); (b) $L_s=75°$ (sol 150); (c) $L_s=180°$ (sol 362). Panels (a)-(b) calculated for climatology average solar and (c) for dust storm average solar scenarios. The wind velocity scale is at right in ms$^{-1}$. Note the change in the wind velocity scale in the different panels. From MCD model.*

Along the meridian circle, the distance from parallels 70°N to 60°N, where the disturbances evolve, to the Perseverance latitude 18.5°N is ~ 3050 - 2450 km. Therefore, pressure disturbances with wavelengths $L_{Bclin}$ ~ 3900 - 5500 km can leave their imprint on the pressure measurements at Perseverance, as observed. However, we note that the size of the disturbances revealed by dust and clouds is smaller, $L$ ~ 300 - 1000 km (in general more elongated meridionally than zonally) (Clancy et al., 2017).

Baroclinic features that appear as spiral disturbances (Hunt and James, 1979) or the annular double cyclone (Sánchez-Lavega et al., 2018) are in gradient wind balance, and tangential rotation velocity $V_T$ is related to the pressure gradient by

$$\frac{V_T^2}{R} + fV_T = -\frac{1}{\rho}\frac{dP}{dr} \tag{1}$$

Measurements of the displacements of clouds and dust masses give tangential velocities of $V_T$ ~ 25 ms$^{-1}$. For $R$ ~ 500 km and density $\rho$ = 0.02 kg m$^{-3}$ ($P$ = 750 Pa, $T$ = 200 K) equation (1) gives $\partial P/\partial r$ ~ 9 Pa/100 km. This crude estimation of the pressure gradient is closer to the actual value in places where the disturbance is more intense and is made visible by the aerosols (most cases dust lifted from the ground). Out of that region, the pressure disturbance should be smaller and according to a geostrophic balance (excluding centrifugal term in (1)), Leovy (1979) and Barnes (1980, 1981) proposed that the measured pressure disturbance at surface is related to the meridional component of the geostrophic wind velocity as

$$v_g \approx \frac{R_g^* T}{f R_M \cos\varphi} \frac{\partial}{\partial \lambda_B}\left(\frac{\delta P}{P_0}\right) = -\frac{R_g^* T}{f c_X} \frac{\partial}{\partial t}\left(\frac{\delta P}{P_0}\right) \tag{2}$$

being $\lambda_B$ the longitude and $c_x$ the zonal phase speed. Barnes (1981) showed that this expression gives probably an overestimate by a factor of 2 when comparing to the measured meridional component of the surface wind speed $v_0$. In our case, we use $c_x$ = 20 ms$^{-1}$ as calculated before and $\delta P$ ~ 2 – 10 Pa (from Figure 10) to determine the expected fluctuation of the wind velocity at Jezero. We find $v_g$ ~ 1 – 3 ms$^{-1}$ at the detection limit of MEDA wind sensors (Newman et al., 2022).

## 6. Short period (minutes) oscillations: gravity waves

The cadence of MEDA measurements along a sol include typically intervals of 1 hr and 5 minutes of data followed by 55 minutes without data (Figure 2). To analyze oscillations within these intervals, we detrended pressure measurements using polynomial fits of different degrees (typically 1- 3). We find that during nighttime the pressure shows regular oscillations that on average have peak to peak amplitudes of 0.2-0.4 Pa and periods in the range between 8 and 24 minutes Figure 11.

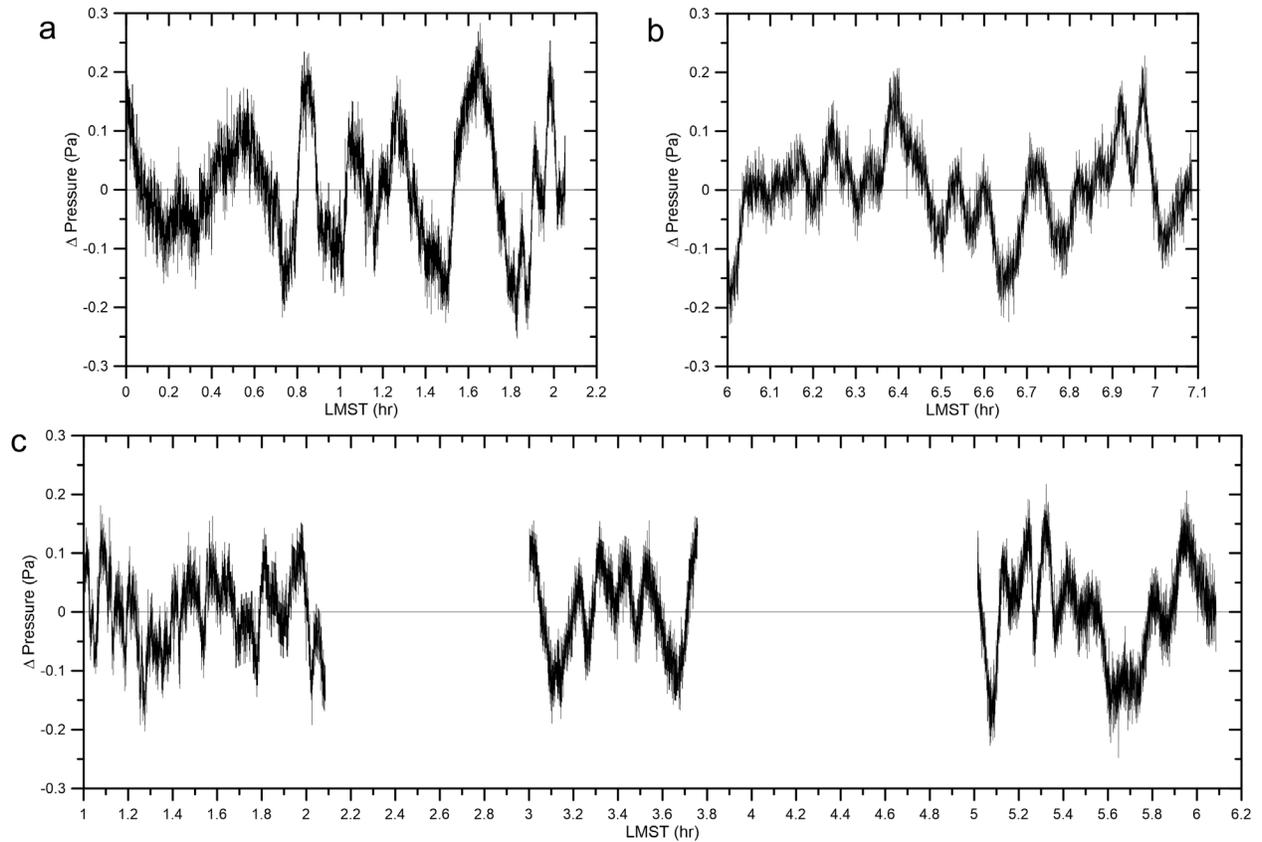

**Figure 11**. *Examples of short period oscillations (12-24 minutes) observed from de-trended residuals in 1-2 hour series in nighttime for sols: (a) 18 (Ls= 14°); (b) 120 (Ls= 62°); (c) 179 (Ls= 88°). The oscillations are obtained from polynomial fits with degrees n=1-3 to the measured pressure data in selected time intervals of ~ 1 hour from MEDA cadence.* In all these cases the coefficient of determination (R-squared) > 0.98 for polynomial terms with degree greater than zero.

During the convective hours the pressure fluctuations are more irregular in periodicity although they usually have shorter periods ~ 6 – 10 min and larger amplitudes, typically in the range of ± 0.4 Pa (Figure 12) in agreement with those found at Gale crater by MSL (Guzewich et al., 2021). Daytime pressure fluctuations also include the short transient pressure drops (ΔP ~ 0.5 – 7 Pa; duration ~ 5 -50 s) due to the close passage of vortices and dust devils (Figure 12) (Newman et al., 2022; Hueso et al., 2022). This daytime activity is due to the development of the convective instability during the maximum insolation hours at Jezero, when the static stability of the atmosphere becomes negative. Newman et al. (2022) have shown that the pressure oscillations are accompanied by temperature and wind velocity oscillations and suggest they are produced by the passage of convection cells (updrafts at cell walls and downdrafts at cell center) with a width of ~ 3-5 km, advected by the large-scale daytime dominant upslope winds at Jezero.

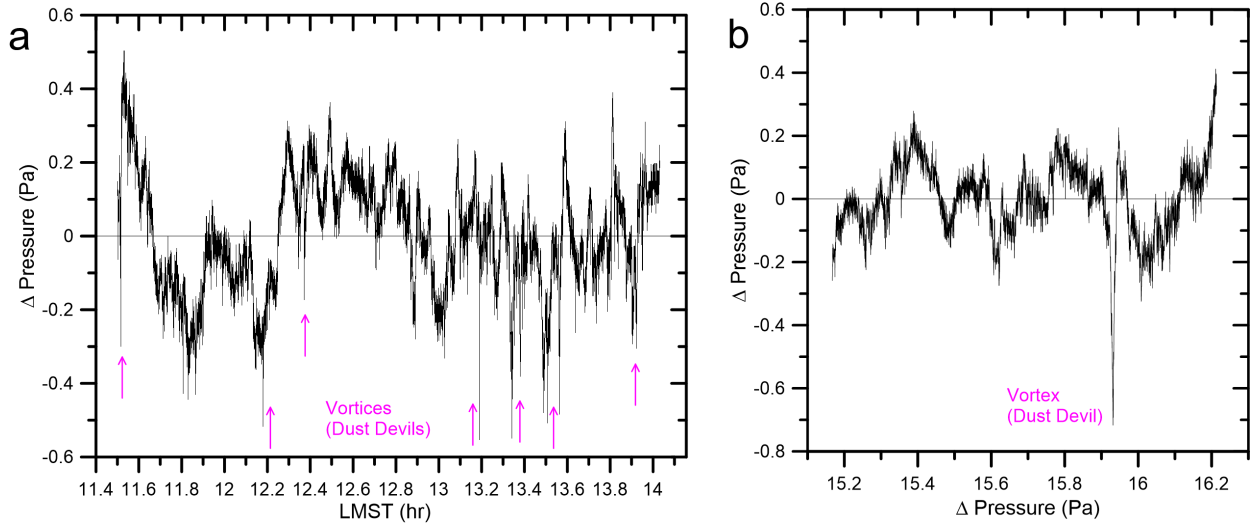

**Figure 12**. *Example of short period fluctuations observed from de-trended residuals in 1-2 hour series in daytime during maximum insolation (convective period) for sol 18 (Ls= 14.5) that includes a series of rapid and deep pressure drops due to the close passage of vortices.*

Nighttime short-period oscillations similar to those shown in Figure 11 were also reported at Gale crater, as observed by the PS on rover Curiosity (Harri et al., 2014), and in Elysium Planitia by the Insight platform (Banfield et al., 2020). Using a simple shallow water model, the oscillations observed at Gale have been interpreted as produced by internal gravity waves excited by cold slope flows in the evening along the walls and central peak of Gale crater (Haberle et al., 2014). We could expect similar excitation mechanism in Jezero as in Gale, but the craters properties are different. Jezero is a shallower crater, about 300 m in depth and 45 km in diameter and Perseverance is close to the interior western rim of the crater, which itself sits on the interior northwest slopes of the ~1350-km-wide Isidis basin. At Jezero, night winds blow from the west-northwest, downslope due to both the Isidis basin and Jezero crater slopes, with low velocities ~ 2-4 ms$^{-1}$ (Newman et al., 2022).

The MEDA temperature sensors allow the retrieval of the temperature gradient from the surface up to an altitude of about 40 m (Rodriguez-Manfredi et al., 2021; Munguira et al., 2022), and the static stability and Brunt-Väisälä frequency of the atmospheric surface layer. The measured vertical temperature gradient between the surface and the 40 m altitude at nighttime for the sol 179 shown in Figure 11 was $dT/dz \sim +0.2$ Km$^{-1}$ (Munguira et al., 2022). Following the MCD model, this gradient decreases to $dT/dz \sim +0.05$ Km$^{-1}$ at the top of the crater. The temperature at LTST 6-7 hr is 195 K and the Brunt-Väisälä frequency corresponding to both gradients is $N_1 = 0.06$ s$^{-1}$ in the ground and $N_2 = 0.032$ s$^{-1}$ at the top of the crater. The observed oscillation periods (~ 8-24 min) correspond to frequencies $\omega \sim 0.013$ s$^{-1}$ – $0.044$ s$^{-1}$.

We study the behavior of internal gravity waves (GW) considering buoyancy as the main driver of the oscillation, i.e. disregarding compression or acoustic terms (Salby, 1996). We also do not consider the effects of the Coriolis force (i.e. inertia-GW) due to the high frequency of the waves and the proximity of the rover to the equator. Under the above hypothesis, the dispersion relationship is given by

$$\omega^* = \omega - uk = \frac{\pm Nk}{\sqrt{k^2 + m^2}} \qquad (3)$$

with $\omega^*$ the intrinsic frequency, $u$ the nighttime wind velocity and the horizontal and vertical wavelengths of the waves given by $k = 2\pi/L_x$ and $m = 2\pi/L_z$. For $m > 0$ the waves propagate upward. We explore three cases for the horizontal wavelength: $L_x = u \cdot \tau \sim 1.9$ and 5.8 km (using $u = 4$ ms$^{-1}$ and $\tau$ the observed oscillation periods) and $L_x = 45$ km (the crater diameter) as an upper limit. The vertical wavelength of these waves as a function of their frequency are shown in Figure 13. The analysis suggests that the range of periods observed for these nocturnal oscillations is compatible with internal gravity waves excited as air flows along the walls of Jezero. Inertia-GW were reported at Gale with much longer horizontal wavelengths ($\sim$ 100-1000 km) (Guzewich et al., 2021).

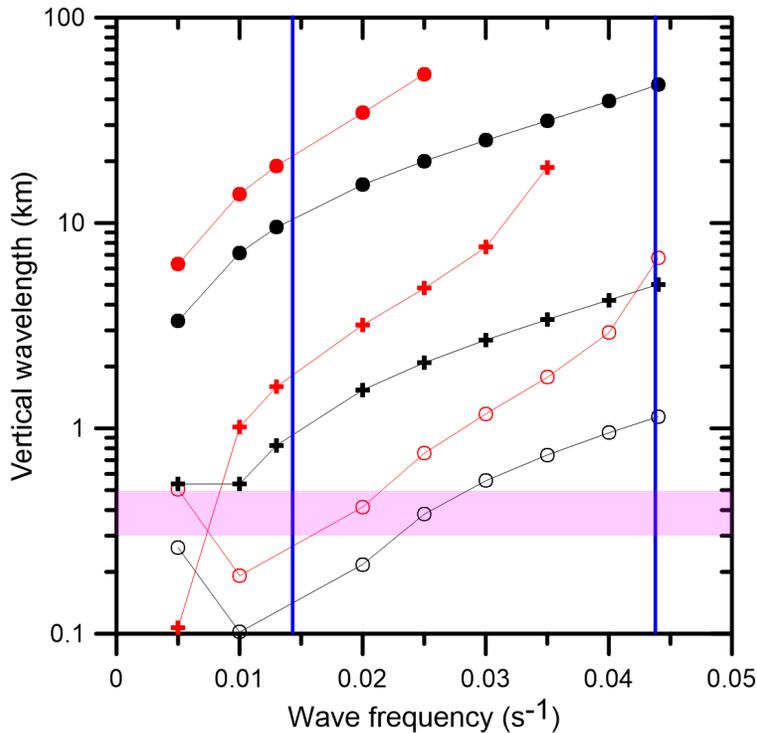

**Figure 13.** *Vertical wavelength of internal gravity waves excited at crater Jezero. The different cases correspond to Brunt-Väisälä frequency $N_1=0.06$ s$^{-1}$(black symbols) and $N_2 = 0.032$ s$^{-1}$ (red symbols). Explored horizontal wavelengths are: $L_x = 1.9$ (dots), $L_x = 5.8$ km (crosses), $L_x = 45$ km (circles). The vertical blue lines mark the typical observed wave periods (24 min and 8 min). The horizontal purple band corresponds to the crater depth.*

If we assume that the vertical wavelength of the internal gravity waves is controlled by the crater depth, then for the Brunt-Väisälä frequency in the range 0.032 s$^{-1}$ - 0.06 s$^{-1}$ as determined from the surface temperature gradients, the corresponding horizontal wavelength is about 2 km (Figure 13), that is, much smaller than the crater diameter.

## 7. Rapid pressure fluctuations (seconds): turbulence

In order to analyse short-time fluctuations in comparison with Banfield et al (2020), we performed 50 second running averages of all intervals having more than 30 minutes of continuous data, and subtracted the pressure signal from them. We observe a clear difference between night and day, with most of night-time fluctuations probably related to detector noise, while a clear signal is always present at daytime. In order to quantify this difference, we calculated 100 bin-histograms of fluctuations up to 1 Pa in two Martian-hour ranges: a "night" range from 0:00 to 6:00 and from 18:00 to 24:00 and a "day" range from 6:00 to 18:00, ignoring the actual daily variations of sunrise and sunset. The histograms were fit to Gaussians to determine half width at half maximum (HWHM), resulting in values of the HWHM with an error of ~ 0.3%. In Figure 14 we present the collective results for the non-dusty (sols 50-280, $L_s$= 30°-136°) and dusty (sols 280-450, $L_s$=136°-235°) seasons. At nighttime, the fit is narrower, leading to Gaussian distribution with amplitude (HWHM) of 0.033 Pa with little dispersion at the tails. However, during the convective period, the Gaussian becomes wider (HWHM amplitude of 0.04 Pa), and the fit is not so good at the tails, with a noticeable increase of fluctuations over 0.07Pa.

Figure 14 also shows sol-by-sol evolution of day and night HWHM of the rapid fluctuations from sols 50 to 455. The mean value of the fluctuations during nighttime hours can be taken at most times as the noise limit in pressure measurements. Note that during the regional dust storm (sols 312-315, $L_s$=152.5°-154.5°) fluctuations increase to levels similar to those observed during daytime hours (see details in Figure 17). The diurnal values also clearly show a temporal uniformity in the HWHM value, although a decrease is observed with the increase in the amount of dust in the atmosphere from sol 350 ($L_s$=174°) onwards.

Such rapid fluctuations have also been reported elsewhere (Spiga et al., 2020; Chatain et al., 2021) and are clearly a result of thermal turbulence generated by convection during maximum daytime heating.

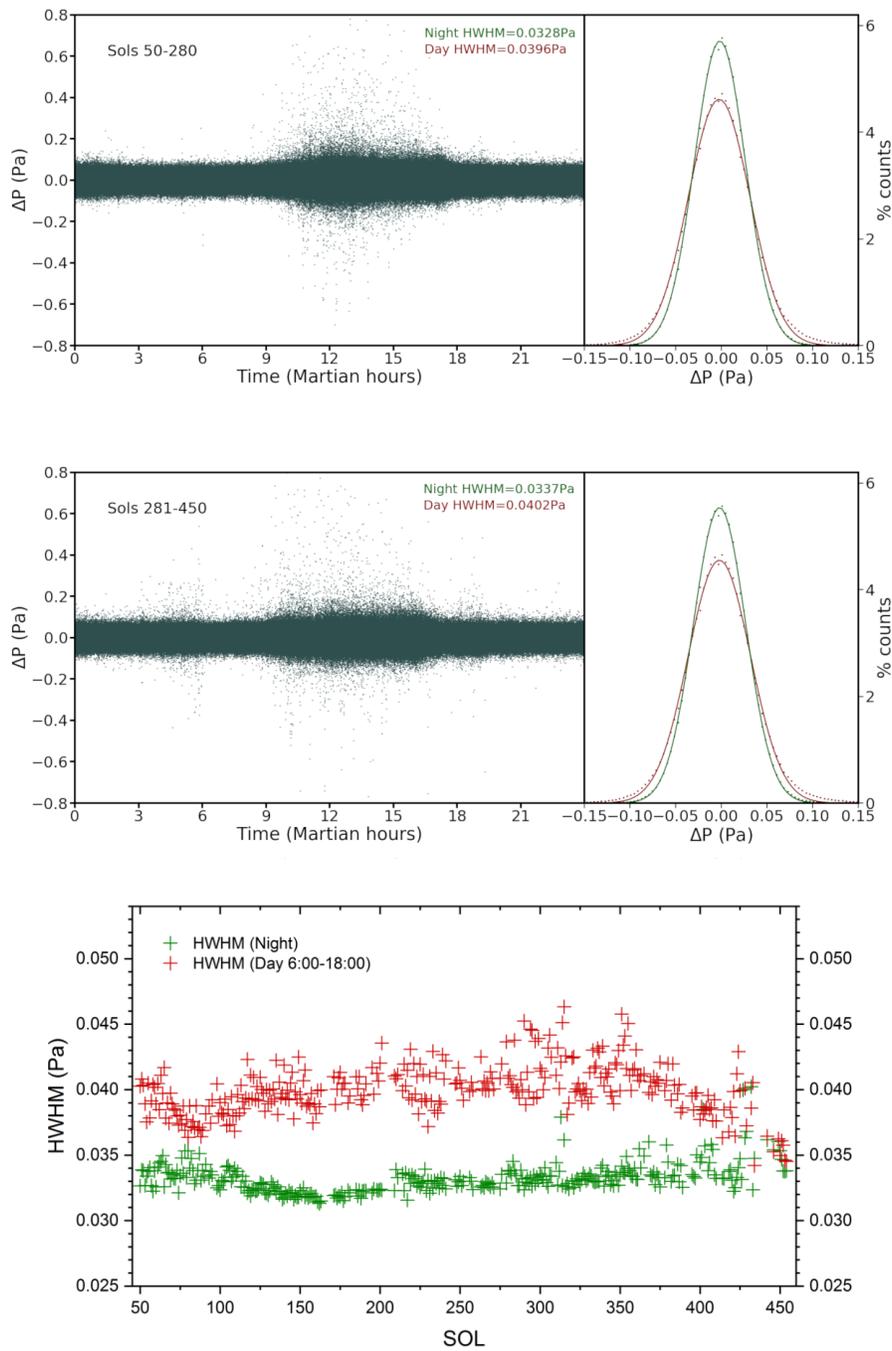

**Figure 14**. *Upper panel: Pressure fluctuations relative to a mean value measured at a frequency of 1 Hz accumulated on a diary basis from measurements between sol 50 and 280 (Ls = 30° – 136°). At right a Gaussian fit corresponding to measurements in daytime (red) and nighttime (green). Middle panel: As before but for sols 281 and 450 (Ls = 136° - 235°). The lower panel shows the integrated view from sols 50 to 455 ($L_s$=30°-238°) of the evolution of the HWHM of the rapid pressure fluctuations in night (green) and day (red).*

## 8. Impact of a regional dust storm on the pressure field

A regional dust storm reached Jezero between 5 and 11 January 2022 [Lemmon et al., 2022] corresponding to sols ~ 312-317 ($L_s$ ~ 155°). The daily pressure cycle was greatly disturbed, with a prominent pressure drop of the daily minimum that reached a peak of ~ 60 Pa in sol 313 at local time ~ 17 hr with respect to sol 311, as shown in Figure 15 (see also Figures 3-5). A similar drop at the same local time but less pronounced (~ 15 Pa) was observed during the decaying stage of a local dust storm over Gale at $L_s$ = 260° in MY32 (Ordoñez-Etxeberria et al., 2020). The global dust storm GDS 2018 at $L_s$ ~ 190° in MY34 (Sánchez-Lavega et al., 2019) showed a more complex behavior in the daily pressure cycle, which was strongly modified at all times of the day (Guzewich et al., 2018; Viudez-Moreiras et al., 2019) due to changes in the tidal components and to the internal circulation in Gale crater. The pressure changes during the 2019 LDS (Large Dust Storm) at $L_s$ ~ 320° in MY34, were analyzed by MSL and at the Insight platform (Viúdez-Moreiras et al., 2020). Similar to the case under study, the diurnal cycle at Insight showed the greatest drop of minimum pressure at the arrival of the storm, of ~ 30 Pa at 17 hr local time, accompanied by a high variability at the other hours of the sol.

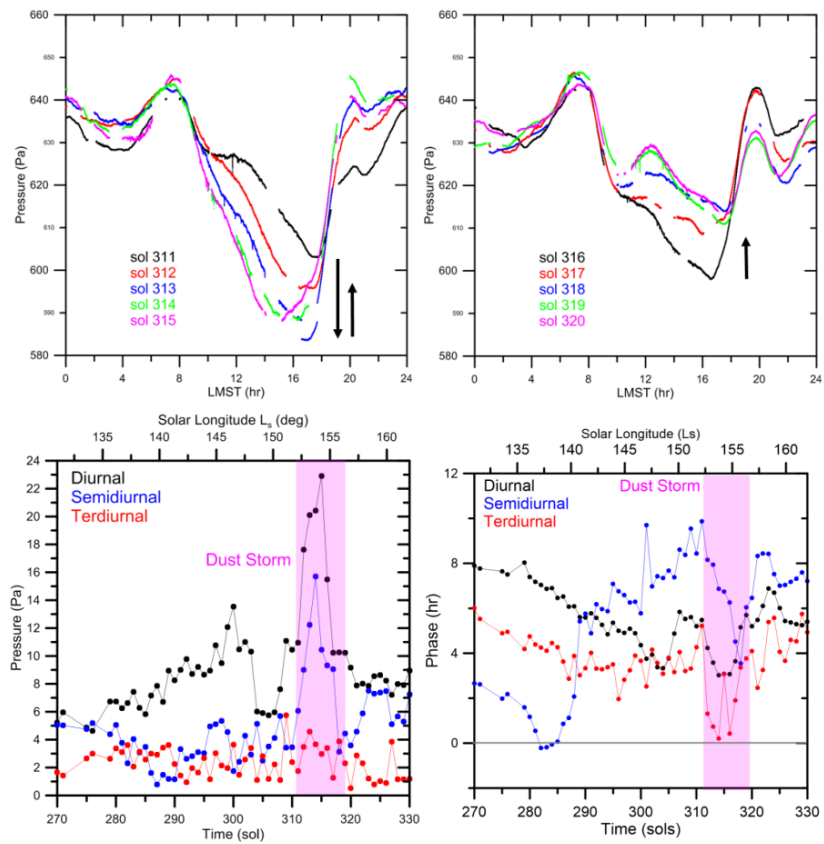

**Figure 15.** *Effects on the pressure field of a regional dust storm that developed over Jezero in early January 2022 ($L_s$ ~ 155°). Upper panels: changes in the daily pressure cycle between sols 311-315 at left and sols 316-320 at right. The arrows marc the trends in the pressure minimum as the storm progressed. Lower panels: changes in the diurnal, semidiurnal and terdiurnal tidal components in amplitude (left) and phase (right).*

The increase in optical depth produced by the storm at Jezero was accompanied by an increase by factors ~ 2.7 and 4 in the amplitude of the diurnal $S_1$ and semidiurnal $S_2$ tidal components (Figure 15) and by a delay of ~ 4 hr and ~ 3 hr in the phase of the diurnal and semidiurnal components. Both phases quickly recovered their previous value following the storm decay. The phase of the terdiurnal tide underwent a similar delay but without a clear central peak increase. The maximum tidal amplitude occurred in sol 313, ~ 3-4 sols in advance of the maximum in optical depth (Figs. 7-8). Simultaneously, the baroclinic wave activity increased its amplitude, with pressure peak-peak-to-peak oscillations reaching ~ 10 Pa (Figure 9). The short period oscillations (section 6) did not ceased during the storm, on the contrary, they became prominent in the morning hours as shown in Figure 16. Peak to peak amplitudes were in the range ~ 0.4 - 0.8 Pa, but short periods in the range 1-3 minutes.

Surface pressure changes associated with the presence of dust storms have been documented since the time of the first landers Viking 1 and 2 (Ryan et al., 1979, 1981). For two global dust storms GDS 1977A and 1977B, both stations measured increases in the amplitude of the diurnal ($S_1$) and semidiurnal ($S_2$) components by a factor 2-6 due to the atmospheric heating produced by the increase in dust optical depth ($\tau$ ~ 3-6) (Zurek, 1981; Wilson and Hamilton, 1996). However, Guzewich et al. (2016) and Ordóñez-Etxeberrí et al. (2019) found a different behavior for the local dust storm in Gale reported above: $S_1$ increased by a factor 1.15 in response to the local opacity enhancement, whereas $S_2$, sensitive to the global averaged dust opacity, exhibited no response. During the arrival of GDS 2018 at MSL, $S_1$ increased by a factor 1.7 and $S_2$ by a factor 3, but no peak was observed in $S_3$ within its general tendency to grow with the increase of dust at that time of the Martian year (Guzewich et al., 2018; Viúdez-Moreiras et al., 2019). These tidal increases in Gale are smaller than those of the storm under study, and smaller than those reported for the 1977 GDSs, again suggesting tidal interactions with the internal crater circulation in Gale. Finally, during the LDS 2019, the optical depth at Insight site increased from $\tau$ ~ 0.7 to 1.9 and $S_1$ and $S_2$ increased by a factor 2 and 1.7 respectively, while the terdiurnal mode $S_3$ increased both at MSL and at Insight (Viúdez-Moreiras et al., 2020). The semidiurnal tide amplitude nicely followed the optical depth path at MSL and Insight sites in good accordance with previous studies. The phase of the diurnal and terdiurnal tides were delayed ~ 2 hr and 1 hr respectively at both sites, but the phase of the semidiurnal mode was earlier by 0.5 hr. The different behavior of thermal tides associated to a variety of storms and at different places, reflects the dependence of thermal tides on dust content and its spatial and temporal evolution. The influence of topography, location on Mars, mutual interactions between modes and other effects probably merit a separate joint analysis of all the cases available.

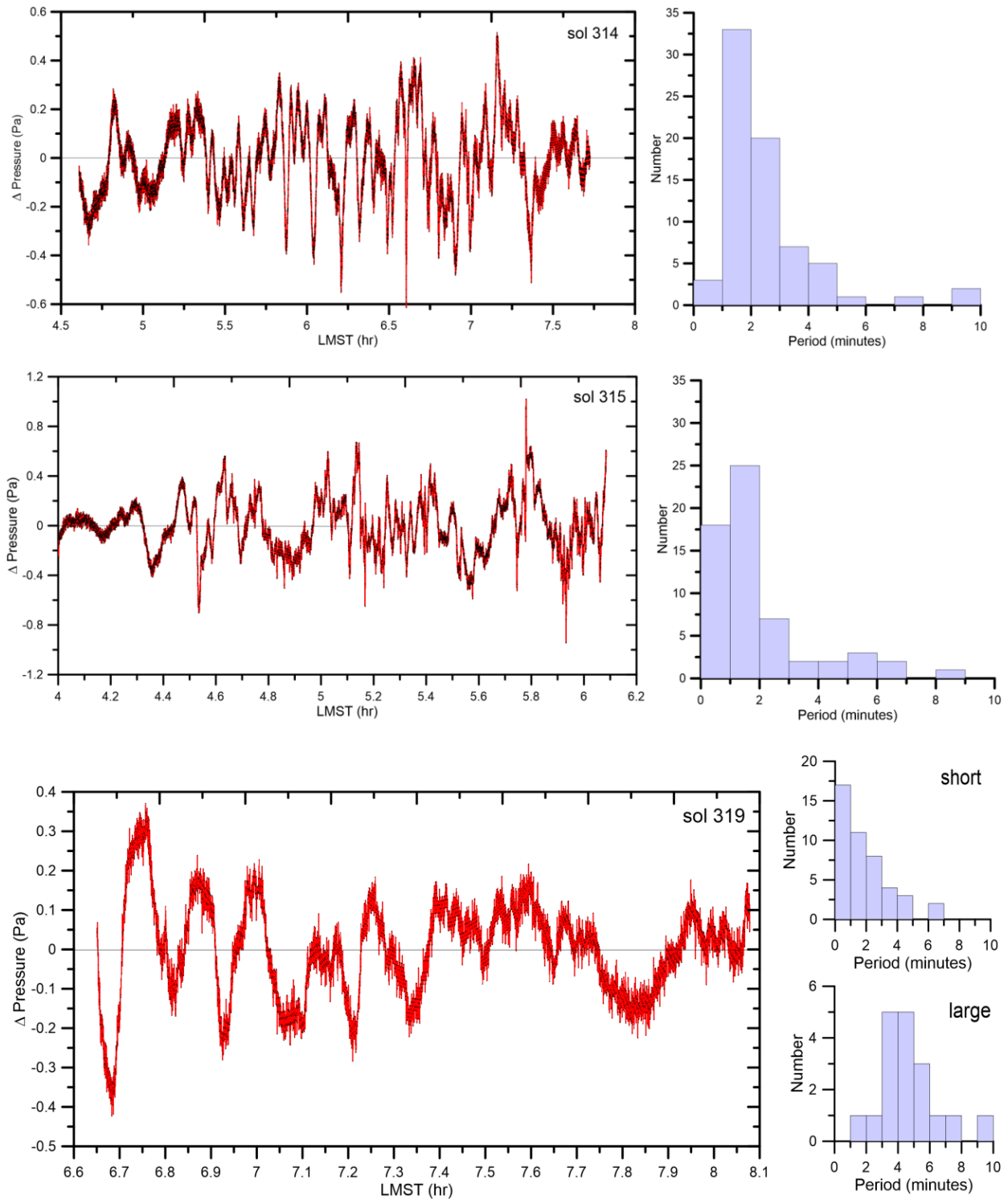

**Figure 16.** *Short period oscillations in the morning hours (4-8 hr LMST) during the evolution of the dust storm in sols 314 (top), 315 (middle) and 319 (bottom). At right are the histograms corresponding to the periodicity of the oscillations.*

Finally, in Figure 17 we present an analysis of the pressure fluctuations similar to that shown in Figure 14 but here concentrating only on the stormy sols (313-315). The rapid fluctuations in the convective hours, with a HWHM of 0.0443 Pa, have increased their amplitude when comparing with the normal situation (Figure 14). Most significantly, rapid fluctuations larger than the noise level appeared also at night-time, increasing the overall nighttime HWHM to 0.0357 Pa.

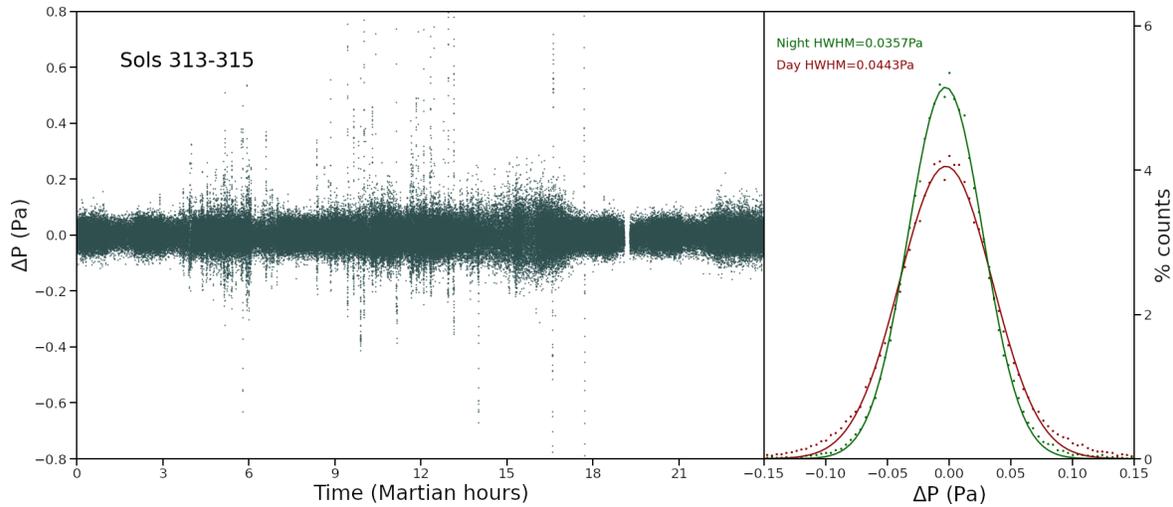

**Figure 17.** *Rapid fluctuations in pressure during the pre-storm and stormy sols 310 - 320 (left panel) and HWHM during night-time (green curve) and daytime (red curve) (right panel).*

## 9. Conclusions

In this paper we have presented a first general analysis of the pressure measurements made by the Perseverance rover between sols 15 and 460 (Ls ~ 13° - 241°) in order to characterize the different atmospheric phenomena involved. Specifically we have found that:

- The seasonal evolution of the mean daily pressure, driven by the CO2 condensation and evaporation cycle at the poles, that compares well with predictions by the MCD model with maximum deviation by about 10 Pa at Ls 145°.

- The amplitude and phase of the first six components of the thermal tides show clear seasonal trends. The maximum value of the amplitudes range from 2 to 22 Pa. The comparison with MCD predictions shows reasonable agreement except for the terdiurnal component.

- We show the correlation between the amplitude of the diurnal and semidiurnal tides with the dust content in the atmosphere, and that a linear combination of the amplitude of the four components is related to the integrated optical depth.

- Long-period waves have been present permanently with periods of 2-5 sols and amplitudes highly variable, between 2 and 10 Pa. The waves amplified with the dust content in the atmosphere. We use the MCD model to interpret this phenomenon as due to baroclinic waves resulting from the instability of the polar jet at the edge of the North Polar Cap. Waves with longer periods are also present in the data.

- Short period oscillations in the range 8-24 minutes and with peak amplitudes ~ 0.4 Pa have been observed regularly at night-time during stable atmospheric conditions. We interpret them as internal gravity waves for Brunt-Väisälä frequencies in the range $0.032$ s$^{-1}$ - $0.06$ s$^{-1}$ (as determined from the surface temperature gradient) and vertical wavelengths of the order of the crater depth. Their expected horizontal wavelength is ~ 2 km (<< Jezero crater diameter).

- In daytime we observed rapid and irregular fluctuations with HWHM amplitudes of 0.04 Pa and maximum peak amplitudes ~ 0.4-0.8 Pa produced by turbulent convection. They are accompanied by transient pressure drops up to ~ 7 Pa with duration ~ 1-150 s due to the close encounter vortices and dust devils.

- We also report the effects on the pressure field produced by a regional dust storm that developed in early January 2022 (sols 310-320). The dust injection produced a maximum drop in the pressure minimum down to 60 Pa in sol 313, accompanied by an increase in the amplitude of the diurnal and semidiurnal tides by a factor of 2-3 together with a drop in their phase. The gravity wave activity in morning hours exhibited a shortening in their periods to 1-3 minutes, and high frequency oscillations above the noise level appeared at nighttime.

**Acknowledgments**

The UPV/EHU team (Spain) is supported by Grant PID2019-109467GB-I00 funded by 1042 MCIN/AEI/10.13039/501100011033/ and by Grupos Gobierno Vasco IT1742-22. GM wants to ac- knowledge JPL funding from USRA Contract Number 1638782. A. Vicente-Retortillo is supported by the Spanish State Research Agency (AEI) Project No. MDM-2017-0737 Unidad de Excelencia "María de Maeztu"- Centro de Astrobiología (INTA-CSIC). Part of the research was carried out at the Jet Propulsion Laboratory, California Institute of Technology, under a contract with the National Aeronautics and Space Administration (80NM0018D0004). GM wants to ac- knowledge JPL funding from USRA Contract Number 1638782.

**Data availability**

MEDA pressure calibrated measurements can be found:
J.A. Rodriguez-Manfredi et al. (2021b)

Derived data produced and reported in this study can be found in:
Sanchez-Lavega, Agustin; del Río-Gaztelurrutia, Teresa; Hueso, Ricardo (2022)

# References

Banfield D., A. Spiga, C. Newman, F. Forget, M. Lemmon, R. Lorenz, N. Murdoch, D. Viudez-Moreiras, J. Pla-Garcia, R. F. Garcia, P. Lognonné, Ö. Karatekin, C. Perrin, L. Martire, N. Teanby, B. V. Hove, J. N. Maki, B. Kenda, N. T. Mueller, S. Rodriguez, T. Kawamura, J. B. McClean, A. E. Stott, C. Charalambous, E. Millour, C. L. Johnson, A. Mittelholz, A. Määttänen, S. R. Lewis, J. Clinton, S. C. Stähler, S. Ceylan, D. Giardini, T. Warren, W. T. Pike, I. Daubar, M. Golombek, L. Rolland, R. Widmer-Schnidrig, D. Mimoun, É. Beucler, A. Jacob, A. Lucas, M. Baker, V. Ansan, K. Hurst, L. Mora-Sotomayor, S. Navarro, J. Torres, A. Lepinette, A. Molina, M. Marin-Jimenez, J. Gomez-Elvira, V. Peinado, J. A. Rodriguez-Manfredi, B. T. Carcich, S. Sackett, C. T. Russell, T. Spohn, S. E. Smrekar, W. B. Banerdt (2020), The atmosphere of Mars as observed by InSight. Nature Geoscience 13, 190–198.

Barnes,J .R. (1980), Time spectral analysis of midlatitude disturbances in the Martian atmosphere, J. Atmos. Sci., 37, 2002-2015.

Barnes J. .R. (1981), Midlatitude disturbances in the Martian atmosphere: A second Mars year, J. Atmos. Sci., 38, 225-234.

Barnes,J .R. (1984), Linear baroclinic instability in the Martian atmosphere, J. Atmos. Sci., 41, 1536-1550.

Barnes, J. R., R. M. Haberle, R. J. Wilson, S. R. Lewis, J. R. Murphy, P. L. Read (2017), The Global Circulation, in The Atmosphere and Climate of Mars (edts. R. M. Haberle, R. Clancy, F. Forget, M. D. Smith and R. W. Zurek), Chapter 9, pp. 228-294, Cambridge University Press, Cambridge, U.K.

Cantor, B. A., James, P. B., Caplinger, M., and Wolff, M. J. (2001), Martian dust storms: 1999 Mars Orbiter Camera observations, J. Geophys. Res. Planets 106 (E10), 23653– 23687. https://doi.org/10.1029/2000JE001310

Chapman S., & R. S. Lindzen (1970), Atmospheric Tides, Thermal and Gravitational, D. Reidel Publishing Co. (Dordrecht, Holland).

Chatain, A., Spiga, A., Banfield, D., Forget, F., & Murdoch, N. (2021). Seasonal variability of the daytime and nighttime atmospheric turbulence experienced by InSight on Mars. Geophysical Research Letters, 48, e2021GL095453, https://doi.org/10.1029/2021GL095453

Clancy, R.T., Montmessin, F., Benson, J., Daerden, F., Colaprete, A., Wolff, M.J. (2017), Mars Clouds. Chapter 5 of The Atmosphere and Climate of Mars, Edited by Haberle, R.M., Clancy, R.T., Forget, F., Simth, M.D. and Zurek, R.W. Cambridge University Press, Cambridge, UK. https://doi.org/10.1017/9781139060172.005

Forget, F., Hourdin, F., Foumier, R., Hourdin, C., Talagrand, O., Collins, M., Lewis, S. R., Read, P.L. (1999), Improved general circulation models of the Martian atmosphere from the surface to above 80 km. Journal of Geophysical Research: Planets, 104(E10), 24155–24175
https://doi: 10.1029/1999JE001025

Guzewich, S. D., Newman, C. E., de la Torre Juarez, M., Wilson, R. J., Lemmon, M., Smith, M. D., et al. (2016). Atmospheric tides in Gale crater, Mars. *Icarus*, *268*, 37–49. https://doi.org/10.1016/j.icarus.2015.12.028

Guzewich, S. D., Lemmon, M., Smith, C. L., Martínez, G., Vicente‐Retortillo, Á., Newman, C. E., et al. (2019). Mars Science Laboratory observations of the 2018/Mars year 34 global dust storm. Geophysical Research Letters, 46, 71–79. https://doi.org/10.1029/2018GL080839

Guzewich, S. D., de la Torre Juárez, M., Newman, C. E., Mason, E., Smith, M. D., Miller, N., et al. (2021). Gravity wave observations by the Mars Science Laboratory REMS pressure sensor and comparison with mesoscale atmospheric modeling with MarsWRF. Journal of Geophysical Research: Planets, 126, e2021JE006907. https://doi.org/10.1029/2021JE006907

Haberle R. M., J. Gómez-Elvira, M. de la Torre Juárez, A.-M. Harri, J. L. Hollingsworth, H. Kahanpää, M. A. Kahre, M. Lemmon, F. J. Martín-Torres, M. Mischna, J. E. Moores, C. Newman, S. C. R. Rafkin, N. Rennó, M. I. Richardson, J. A. Rodríguez-Manfredi, A. R. Vasavada, M.-P. Zorzano-Mier, REMS/MSL Science Teams (2014), Preliminary interpretation of the REMS pressure data from the first 100 sols of the MSL mission, J. Geophys. Res. Planets, 119, 440– 453, https://doi:10.1002/2013JE004488

Haberle, R.M., de la Torre Juárez, M., Kahre, M.A., Kass, D.M., Barnes, J.R., Hollingsworth, J.L., Harri, A.-M., Kahanpää, H. (2018), Detection of Northern Hemisphere transient eddies at Gale Crater Mars, Icarus, 307, 150-160.

Harri A.-M., M. Genzer, O. Kemppinen, H. Kahanpää, J. Gomez-Elvira, J. A. Rodriguez-Manfredi, R. Haberle, J. Polkko, W. Schmidt, H. Savijärvi, J. Kauhanen, E. Atlaskin, M. Richardson, T. Siili, M. Paton, M. de la Torre Juarez, C. Newman, S. Rafkin, M. T. Lemmon, M. Mischna, S. Merikallio, H. Haukka, J. Martin-Torres, M.-P. Zorzano, V. Peinado, R. Urqui, A. Lapinette, A. Scodary, T. Mäkinen, L. Vazquez, N. Rennó; the REMS/MSL Science Team (2014), Pressure observations by the Curiosity rover: Initial results. J. Geophys. Res. Planets 119*,* 2132–2147. https://doi: 10.1002/2013JE004423

Harri, A.-M, M. Paton, M. Hieta, J. Polkko, C. E. Newman, J. Pla-García, J. Leino, J. Kauhanen, I. Jaakonaho, R. Hueso, A. Sánchez-Lavega, M. Genzer, R. Lorenz, M. Lemmon, A. Vicente-Retortillo, L. K. Tamppari, D. Viudez-Moreiras, M. de la Torre-Juarez, H. Savijärvi, J. A. Rodríguez-Manfredi, G. Martinez (2022), Perseverance MEDA-PS pressure observations – initial results (this issue)

Hess, S.L., Henry, R.M., Levoy, C.B., Ryan, J.A., Tillman, J.E. (1977), Meteorological results from the surface of Mars: Viking 1 and 2, Journal of Geophysical Research, 82, 4559-4574.

Hess S.L., J.A. Ryan, J.E. Tillman, R.M. Henry, C.B. Leovy, The annual cycle of pressure on Mars measured by Viking landers 1 and 2. Geophys. Res. Lett. 7(3), 197–200 (1980)

Hinson David P., Huiqun Wang (2010), Further observations of regional dust storms and baroclinic eddiesin the northern hemisphere of Mars, Icarus 206, 290–305
https://www.sciencedirect.com/science/article/pii/S0019103509003613

Hueso, R., A. Munguira, A. Sánchez-Lavega, C. E. Newman, M. Lemmon, T. del Río-Gaztelurrutia, M. Richardson, V. Apestigue, D. Toledo, A. Vicente-Retortillo, M. de la Torre-Juarez, J. A. Rodríguez-Manfredi, L. K. Tamppari, I. Arruego, N. Murdoch, G. Martinez, S. Navarro, J. Gómez-Elvira, M. Baker, R. Lorenz, J. Pla-García, A.M. Harri, M. Hieta, M. Genzer, J. Polkko, I. Jaakonaho, T. Mäkinen, A. Stott, D. Mimoun, B. Chide, E. Sebastian, D. Viudez-Moreiras, D. Banfield, A. Lepinette-Malvite (2022), Vortex and dust devil activity on jezero crater from Mars2020/meda data and physical characterization of selected events, Mars Atmospheric Modelling and Observations, 7th Workshop, 14-17 June 2022 2020 (Paris, France).

Hunt, G. E., and James, P. B., 1979. Martian extratropical cyclones. Nature 278, 531– 532.
https://doi.org/10.1038/278531a0

Khare , M.A., Murphy, J.R., Newman, C.E., Wilson, R.J., Cantor, B.A., Lemmon, M.T., Wolff, M.J., 2017. The Mars Dust Cycle, chapter 10 of The Atmosphere and Climate of Mars, Edited by Haberle, R.M., Clancy, R.T., Forget, F., Smith, M.D. and Zurek, R.W. Cambridge University Press, Cambridge, UK., https://doi.org/10.1017/9781139060172.010

Lemmon M.T., M.D. Smith, D. Viudez-Moreiras, M. de la Torre-Juarez, A. Vicente-Retortillo, A. Munguira, A. Sanchez-Lavega, R. Hueso, G. Martinez, B. Chide, R. Sullivan, D. Toledo, L. Tamppari, T. Bertrand, J.F. Bell III, C. Newman, M. Baker, D. Banfield, J.A. Rodriguez-Manfredi, J.N. Maki, V. Apestigue (2022), Dust, Sand, and Winds within an Active Martian Storm in Jezero Crater, Geophys. Res. Lett. (in the press),
https://doi.org/10.1029/2022GL100126

Leovy, C.B. (1979), Martian Meteorology, Ann. Rev. Astron. Astrophys. 17, 387–413.

Millour, E., Forget, F., Spiga, A., Navarro, T., Madeleine, J. -B., Montabone, L., Pottier, A., Lefevre, F., Montmessin, F., Chaufray, J. -Y., Lopez-Valverde, M. A., Gonzalez-Galindo, F., Lewis, S. R., Read, P. L., Huot, J. -P., Desjean, M. -C., MCD/GCM development Team, 2015. The Mars Climate Database (MCD version 5.2). European Planetary Science Congress 2015.

Munguira, A., Hueso, R., Sánchez-Lavega, A., De la Torre-Juarez, M., Martinez, G., Newman C., Pla-García, J., Banfield, D., Vicente-Retortillo, A., Lepinette, A., Rodríguez-Manfredi, J.A., Chide, B., Bertrand, T., Lemmon, M., Sebastian, E., Navarro, S., Gómez-Elvira, J., Torres, J., Martín-Soler, J., Romeral, J., Lorenz, R., (2022). Mars 2020 MEDA Measurements of Near-Surface Atmospheric Temperatures at Jezero, Mars Atmospheric Modelling and Observations, 7th Workshop, 14-17 June 2022 2020 (Paris, France)

Nelder, John A.; R. Mead (1965). A simplex method for function minimization. *Computer Journal*. **7** (4): 308–313. https://doi:10.1093/comjnl/7.4.308

Newman C. E., M. de la Torre Juárez, J. Pla-García, R. J. Wilson, S. R. Lewis, L. Neary, M. A. Kahre, F. Forget, A. Spiga, M. I. Richardson, F. Daerden, T. Bertrand, D. Viúdez-Moreiras, R. Sullivan, A. Sánchez-Lavega, B. Chide, J. A. Rodriguez-Manfredi (2021), Multi-model meteorological and aeolian predictions for Mars 2020 and the Jezero crater region, Space Sci. Rev. **217**, 20.

Newman C., R. Hueso, M. T. Lemmon, A. Munguira, A. Vicente-Retortillo, V. Apestigue, G. Martinez, D. Toledo, R. Sullivan7, K. Herkenhoff8, M. de la Torre-Juarez, M. I. Richardson, A. Stott, N. Murdoch, A. Sánchez-Lavega, M.J. Wolff, I. Arruego, E. Sebastián, S. Navarro, J. Gómez-Elvira, L. Tamppari, D. Viúdez-Moreiras, A.-M. Harri, M. Genzer, M. Hieta, R.D. Lorenz, P. Conrad, F. Gómez, T.H. McConnochie, D. Mimoun, C. Tate, T. Bertrand, J.F. Bell III, J.N. Maki, J. Antonio Rodriguez-Manfredi, R.C. Wiens, B. Chide, S. Maurice, M.-P. Zorzano, L. Mora, M.M. Baker, D. Banfield, J. Pla-Garcia, O. Beyssac, A. Brown, B. Clark,A. Lepinette, F. Montmessin, E. Fischer, P. Patel, T. del Río-Gaztelurrutia, T. Fouchet, R. Francis, S.D. Guzewich (2022). The dynamic atmospheric and aeolian environment of Jezero crater, Mars. (2022), Science Advances, 8, eabn3783 DOI: 10.1126/sciadv.abn3783

Ordoñez-Etxeberria, I., Hueso, R., Sánchez-Lavega, A. (2019), Meteorological pressure at Gale crater from a comparison of REMS/MSL data and MCD modelling: Effect of dust storms, Icarus, Volume 317, 591-609, https://doi.org/10.1016/j.icarus.2018.09.003

Ordoñez-Etxeberria, I., Hueso, R., Sanchez-Lavega, A. y Vicente-Retortillo, A. (2020). Characterization of a local dust storm on Mars with REMS/MSL measurements and MARCI/MRO images. Icarus 338, 113521, 1-19. https://doi.org/10.1016/j.icarus.2019.113521

Pla-García J., S. C. R. Rafkin, G. M. Martinez, Á. Vicente-Retortillo, C. E. Newman, H. Savijärvi, M. de la Torre, J. A. Rodriguez-Manfredi, F. Gómez, A. Molina, D. Viúdez-Moreiras, A. M. Harri (2020), Meteorological predictions for *Mars 2020 Perseverance Rover* landing site at Jezero crater. Space Sci Rev. 216, 148.

Read P.L., S. R. Lewis and D. P. Mulholland (2015), The physics of Martian weather and climate: a review. Rep. Prog. Phys. 78, 125901, 10.1088/0034-4885/78/12/125901

Read P. L., B. Galperin, S. E. Larsen, S. R. Lewis, A. Määttänen, A. Petrosyan, N. Rennó, H. Savijärvi, T. Siili, A. Spiga, A. Toigo, L. Vázquez, The Martian Planetary Boundary Layer (2017), Chapter 7 of The Atmosphere and Climate of Mars, Edited by Haberle, R.M., Clancy, R.T., Forget, F., Simth, M.D. and Zurek, R.W. Cambridge University Press, Cambridge, UK. https://doi.org/10.1017/9781139060172.005

Richardson, M. I., & Newman, C. E. (2018). On the relationship between surface pressure, terrain elevation, and air temperature. Part I: The large diurnal surface pressure range at Gale crater, Mars and its origin due to lateral hydrostatic adjustment. Planetary and Space Science, 164, 132–157. https://doi.org/10.1016/j.pss.2018.07.003

Rodriguez-Manfredi J. A., M. de la Torre Juárez, A. Alonso, V. Apéstigue, I. Arruego, T. Atienza, D. Banfield, J. Boland, M. A. Carrera, L. Castañer, J. Ceballos, H. Chen-Chen, A. Cobos, P. G.

Conrad, E. Cordoba, T. del Río-Gaztelurrutia, A. de Vicente-Retortillo, M. Domínguez-Pumar, S. Espejo, A. G. Fairen, A. Fernández-Palma, R. Ferrándiz, F. Ferri, E. Fischer, A. García-Manchado, M. García-Villadangos, M. Genzer, S. Giménez, J. Gómez-Elvira, F. Gómez, S. D. Guzewich, A.-M. Harri, C. D. Hernández, M. Hieta, R. Hueso, I. Jaakonaho, J. J. Jiménez, V. Jiménez, A. Larman, R. Leiter, A. Lepinette, M. T. Lemmon, G. López, S. N. Madsen, T. Mäkinen, M. Marín, J. Martín-Soler, G. Martínez, A. Molina, L. Mora-Sotomayor, J. F. Moreno-Álvarez, S. Navarro, C. E. Newman, C. Ortega, M. C. Parrondo, V. Peinado, A. Peña, I. Pérez-Grande, S. Pérez-Hoyos, J. Pla-García, J. Polkko, M. Postigo, O. Prieto-Ballesteros, S. C. R. Rafkin, M. Ramos, M. I. Richardson, J. Romeral, C. Romero, K. D. Runyon, A. Saiz-Lopez, A. Sánchez-Lavega, I. Sard, J. T. Schofield, E. Sebastian, M. D. Smith, R. J. Sullivan, L. K. Tamppari, A. D. Thompson, D. Toledo, F. Torrero, J. Torres, R. Urquí, T. Velasco, D. Viúdez-Moreiras, S. Zurita; The MEDA Team, (2021a). The Mars Environmental Dynamics Analyzer, MEDA. A suite of environmental sensors for the Mars 2020 mission. Space Sci. Rev. 217, 1-86, doi: 10.1007/s11214-021-00816-9

Rodriguez-Manfredi J.A. et al. (2021b), The Mars Environmental Dynamics Analyzer, MEDA, NASA Planetary Data System, DOI 10.17189/1522849

Ryan, J. A., & Henry, R.M. (1979). Mars atmospheric phenomena during major dust storms, as measured at surface. Journal of Geophysical Research, 84(B6), 2821–2829. https://doi.org/10.1029/JB084iB06p02821

Ryan, J. A., & Sharmann, R. M. (1981). Two major dust storms, one Mars year apart: Comparison from Viking data. Journal of Geophysical Research, 86(C4), 3247–3254. https://doi.org/10.1029/JC086iC04p03247

Salby, M. L. (1996), Fundamentals of Atmospheric Physics, Volume 61 International Geophysics, Elsevier

Sánchez-Lavega, A., Garro, A., del Río-Gaztelurrutia, T., Hueso, R., Ordoñez-Etxeberria, I., Chen Chen, H., Cardesín-Moinelo, A., Titov, D., Wood, S., Almeida, M., Spiga, A., Forget, F., Määttänen, A., Hoffmann, H., Gondet, B (2018). A seasonally recurrent annular cyclone in Mars northern latitudes and observations of a companion vortex. J. Geophys. Res. Planets 123, 3020–3034. https://doi.org/10.1029/2018JE005740

Sánchez‐Lavega, A., del Río-Gaztelurrutia, T., Hernández‐Bernal, J., & Delcroix, M. (2019). The onset and growth of the 2018 Martian global dust storm. Geophysical Research Letters, 46, 6101–6108. https://doi.org/10.1029/2019GL083207

Sánchez-Lavega A., A. Erkoreka, J. Hernández-Bernal, T. del Río-Gaztelurrutia, J. García-Morales, I. Ordoñez-Etxeberría, A. Cardesín-Moinelo, D. Titov, S. Wood, D. Tirsch, E. Hauber, K.-D. Matz (2022), Cellular patterns and dry convection in textured dust storms at the edge of Mars North Polar Cap, Icarus, 387, 115183. https://doi.org/10.1016/j.icarus.2022.115183

Sánchez-Lavega, Agustin; del Río-Gaztelurrutia, Teresa; Hueso, Ricardo (2022): M2020 Perverance Mars Pressure Analysis. figshare. Dataset. https://doi.org/10.6084/m9.figshare.20938039

Spiga, A., Murdoch, N., Lorenz, R., Forget, F., Newman, C., Rodriguez, S., et al. (2021). A study of daytime convective vortices and turbulence in the Martian planetary boundary layer based on half-a-year of InSight atmospheric measurements and large-eddy simulations. *Journal of Geophysical Research: Planets*, 126, e2020JE006511. https://doi.org/10.1029/2020JE006511

Tyler, D. Jr. & Barnes, J. R. (2005). A mesoscale model study of summertime atmospheric circulations in the north polar region of Mars. J. Geophys. Res. Planets 110, E06007, 1-26. https://doi.org/10.1029/2004JE002356

Tyler, D., Jr., and J. R. Barnes (2015), Convergent crater circulations on Mars: Influence on the surface pressure cycle and the depth of the convective boundary layer, Geophys. Res. Lett., 42, 7343–7350, doi:10.1002/2015GL064957.

Vallis G. K. (2006), Atmospheric and Ocean Fluid Dynamics, Fundamentals and Large-Scale Circulation, Cambridge University Press

Viúdez-Moreiras, D., Newman, C. E., Torre, M., Martínez, G., Guzewich, S., Lemmon, M., et al (2019). Effects of the MY34/2018 Global Dust Storm as measured by MSL REMS in Gale Crater. J. Geophys. Res. Planets, 124. 10.1029/2019JE005985.

Viúdez-Moreiras D., C. E. Newman, F. Forget, M. Lemmon, D. Banfield, A. Spiga, A. Lepinette, J. A. Rodriguez-Manfredi, J. Gómez-Elvira, J. Pla-García, N. Muller, M. Grott (2020), Effects of a large dust storm in the near-surface atmosphere as measured by InSight in Elysium Planitia, Mars. Comparison with contemporaneous measurements by Mars Science Laboratory. J. Geophys. Res. 125, e2020JE006493.

Wang, H., Richardson, M. I., Wilson, R. J., Ingersoll, A. P., Toigo, A. D., and Zurek, R. W. (2003), Cyclones, tides, and the origin of a cross-equatorial dust storm on Mars, Geophys. Res., Lett., 30, 1488, doi:10.1029/2002GL016828

Wang H.Q., R.W. Zurek, M.I. Richardson, Relationship between frontal dust storms and transient eddy activity in the northern hemisphere of Mars as observed by Mars Global Surveyor, J. Geophys. Res. Planets, 110 (2005), p. E07005, 10.1029/2005JE002423

Wilson, R. J. & K. Hamilton (1996), Comprehensive model simulations of thermal tides in the Martian atmosphere, J. Atmos. Sci., 53(9), 1290 – 1326. https://doi.org/10.1175/1520-0469(1996)053<1290:CMSOTT>2.0.CO;2

Wilson, R. J., Lewis, S. R., &Montabone, L. (2008), Thermal tides in an assimilation of three years of Thermal Emission Spectrometer data from Mars Global Surveyor. Third International Workshop on the Mars Atmosphere: Modelling and Observations workshop, Williamsburg, VA.

Wood, S. E., and D. A. Paige (1992),Modeling the Martian seasonal CO2 cycle.2: Interannual variability, Icarus, 99, 1-14.

Zurek R.W. (1976), Diurnal Tide in the Martian Atmosphere, J. Atmos. Sci., 33, 321-337

Zurek, R., Leovy, C.B., (1981), Thermal tides in the dusty martian atmosphere: A verification of theory. Science 213, 437–439.
http://dx.doi.org/10.1126/science.213.4506.437